\documentclass[aps,twocolumn,10pt,prl,superscriptaddress,floatfix,nofootinbib]{revtex4-2}
\usepackage[T1]{fontenc}
\usepackage[latin9]{inputenc}
\usepackage{xcolor}
\usepackage{xspace}
\usepackage{amssymb,amsmath}
\usepackage{amsbsy}
\usepackage{graphicx}
\usepackage{multirow}
\usepackage{rotating}
\usepackage{tabularx}
\usepackage{float}
\usepackage[normalem]{ulem}
\usepackage{tikz}
\usepackage{hyperref}

\def\ket#1{{|#1\rangle}}
\def\bra#1{{\langle#1|}}
\def\ketbra#1#2{\ket{#1}\bra{#2}}

\DeclareMathOperator{\tr}{tr}

\newcommand{\gap}[1]{\omega_{#1} \sigma^z_{#1}}

\newcommand{\opn}[1]{\operatorname{#1}}

\begin{document}
\title{Thermodynamics of  a minimal algorithmic cooling refrigerator}
\author{Rodolfo R. Soldati}
\affiliation{Institute for Theoretical Physics I, University of Stuttgart, D-70550 Stuttgart, Germany}
\author{Durga B. R. Dasari }
\affiliation{3rd Institute of Physics, IQST, and Research Centre SCoPE, University of Stuttgart, Stuttgart, Germany}
\author{J\"{o}rg Wrachtrup}
\affiliation{3rd Institute of Physics, IQST, and Research Centre SCoPE, University of Stuttgart, Stuttgart, Germany}
\affiliation{Max Planck Institute for Solid State Research, Stuttgart, Germany}
\author{Eric Lutz}
\affiliation{Institute for Theoretical Physics I, University of Stuttgart, D-70550 Stuttgart, Germany}

\begin{abstract}
We  investigate, theoretically and experimentally, the thermodynamic performance of a minimal three-qubit heat-bath algorithmic cooling refrigerator. We analytically compute the coefficient of performance, the cooling power and the polarization of the target qubit for an arbitrary number of  cycles, taking realistic experimental imperfections into account. We determine their fundamental upper bounds in the ideal reversible limit and show that these values may be experimentally approached  using a system of three qubits in a nitrogen-vacancy center in diamond.

\end{abstract}

\maketitle

Cooling has been  an important application of  thermodynamics since its foundation. Refrigeration generically occurs when heat is  extracted from a system, leading to a decrease of its entropy and a reduction of its  temperature below that of the environment \cite{cal85}. Efficient cooling methods are essential for the study  of low-temperature quantum  phenomena, from the physics of atoms and molecules \cite{met99,let09} to novel states of matter  \cite{mac92,ens05} and the development of quantum technologies \cite{nie00,des09}. In the latter context, the challenge to initialize qubits in pure states  with high fidelity has led to the introduction of  powerful algorithmic cooling techniques, in which standard quantum logic gates are employed to transfer heat  out of a number of spins in order to increase their polarization, both in  closed \cite{sch99} and open \cite{boy02}   systems (see Ref.~\cite{par16} for a review).

Heat-bath algorithmic cooling  is a method that allows to cool (slow-relaxing) target  spins with the help of (fast-relaxing) reset spins that pump entropy out of the target spins into a  bath,  which acts  as an entropy sink \cite{boy02,par16,fer04,sch05,sch07,rem07,kay07,bra14,rai15,rod16,rai19,rai21}.  An   algorithmic cooling cycle   consists of a succession of (i) compression steps that cool the target spins and heat up the reset spins, and    of (ii) refresh steps during which the reset spins quickly relax back to the bath temperature (Fig.~1).
Cyclic  algorithmic cooling operation has recently been demonstrated experimentally for a minimal system  of three qubits, using nuclear magnetic resonance \cite{bau05,rya08,par15,ata16} and nitrogen-vacancy  centers in diamond \cite{zai21}.

Motivated by these experiments, we here {introduce a realistic model of  a   heat-bath algorithmic cooling refrigerator composed of one target qubit and of two reset qubits \cite{bau05,rya08,par15,ata16,zai21} and investigate   its thermodynamic performance.} We determine its fundamental limits and compare them to those of standard quantum refrigerators \cite{kos14,rez09,all10,aba16,com1}. Conventional refrigerators cyclically pump heat from a cold to a hot macroscopic system (both considered as heat baths) by consuming work  \cite{cal85}. Two central figures of merit of such refrigerators are the coefficient of performance (COP), defined as the ratio of heat extracted and work supplied, and the cooling power    that characterizes the rate of heat removal. The maximum value of the COP is given, in the reversible limit, by the ideal Carnot expression, \(\zeta_\text{C} = T_\text{c}/(T_\text{h}- T_\text{c})\), where \(T_\text{c}\) and \(T_\text{h}\) are the respective temperatures of the cold and hot baths \cite{cal85}. Algorithmic cooling refrigerators share similarities  with conventional quantum refrigerators: they   cyclically transfer heat from  the cold  spins to the hot bath  by consuming    work done  by    gate operations. Such analogy makes a comparison between  the two   refrigerators  possible.   However, their underlying cooling  mechanisms are intrinsically different and the finite size of the target qubit results in a   cycle that is not closed in the thermodynamic sense, since its state is not  the same at the beginning and at the end of one cycle.

\begin{figure}[t]
    \centering
    \includegraphics[width=0.44\textwidth]{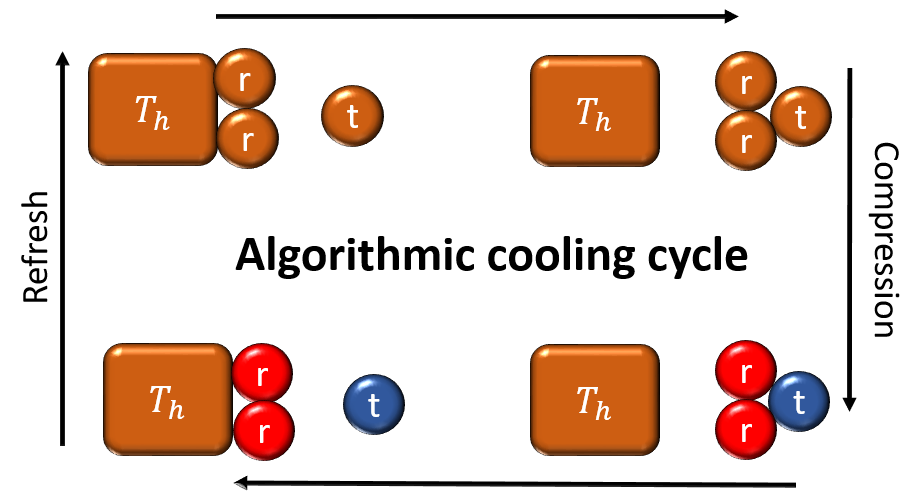}
    \caption{Schematic illustration of the minimal three-qubit algorithmic cooling cycle: in a first (compression) step, heat  is extracted from the target qubit (t), cooling it down while heating up the two reset qubits (r). In a second (refresh) step, the reset qubits are rethermalized to the bath temperature \(T_\text{h}\).}
    \label{fig1}
\end{figure}

\begin{figure*}[t]
	\centering
	\begin{tikzpicture}
        \node (a) [label={[label distance=-.65 cm]145:\textbf{a)}}]  at (-7.2,0) {\includegraphics[width=0.31\textwidth]{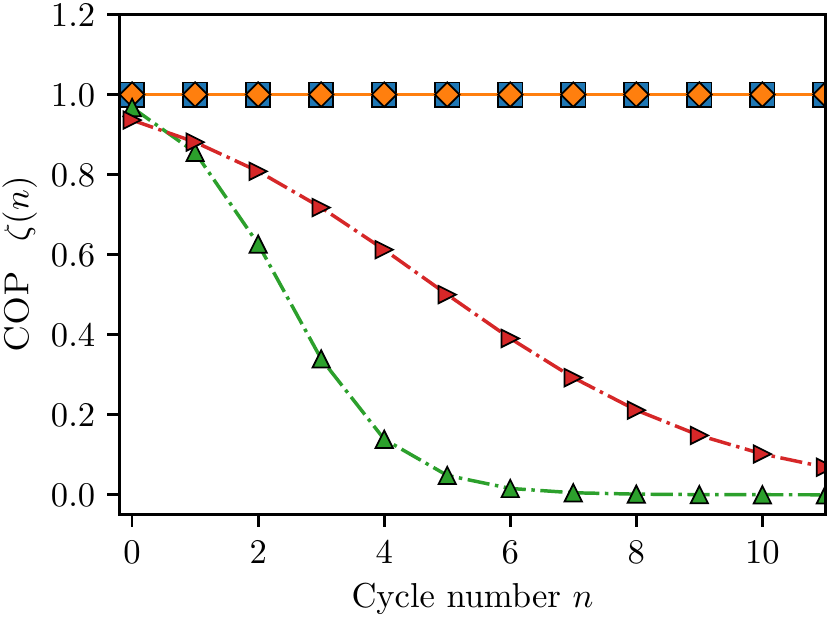}};
        \node (a) [label={[label distance=-.65 cm]145:\textbf{b)}}]  at (-1.2,0) {\includegraphics[width=0.31\textwidth]{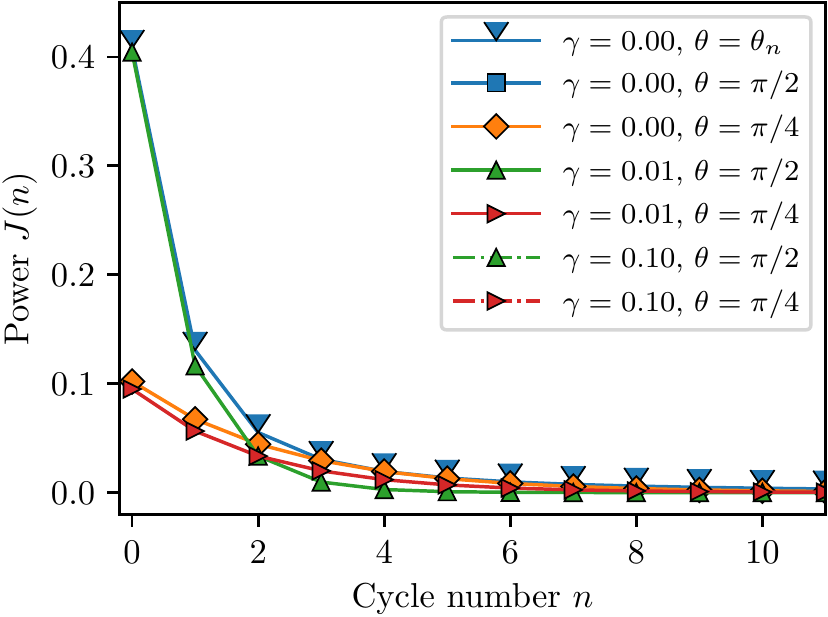}};
        \node (a) [label={[label distance=-.65 cm]145:\textbf{c)}}]  at (4.8,0) {\includegraphics[width=0.31\textwidth]{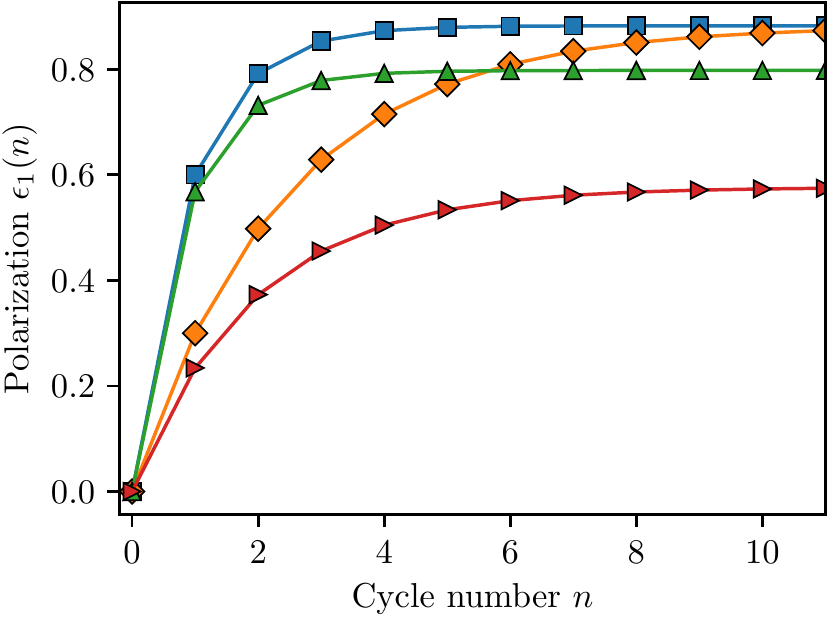}};
	\end{tikzpicture}
	\caption{Thermodynamic performance of the algorithmic cooling refrigerator per cycle. a) Coefficient of performance  \(\zeta(n)\), Eq.~(8),  b) Cooling power \(J(n)\), Eq.~(8), and c) Polarization of the target qubit \(\epsilon_1(n)\), Eq.~(9), for various values of the damping rate \(\gamma\) and of the mixing angle \(\theta\). These two parameters have radically different effects:  whereas the decay constant affects the asymptotic value of the polarization, the mixing angle changes the convergence rate to that value. In addition, the behavior of the cooling power mostly depends of the mixing angle, while the COP depends on both variables. The fundamental upper bounds in the reversible limit {\((\gamma=0\))} are shown by the blue squares. {Parameters are \(\epsilon_1(0)=0\), \(\epsilon_2(0)=\epsilon_3(0)= \epsilon=0.6\).}}\label{fig:appfig1}
\end{figure*}

The performance of thermal machines coupled to finite baths  with finite heat capacities may be conveniently analyzed with  cycle-dependent quantities \cite{ond81,wan16,taj17,poz18,moh19,ma20}. In the following, we compute  COP,  cooling power and  polarization  of the target qubit per cycle for an arbitrary number of cycle iterations. We employ Liouville space techniques \cite{gya20} to exactly solve the full nonstationary dynamics of the system. While heat-bath algorithmic cooling has been mostly studied in the unitary  limit and under steady-state conditions \cite{boy02,par16,fer04,sch05,sch07,rem07,kay07,bra14,rai15,rod16,rai19,rai21}, we explicitly account for experimentally relevant external damping of the target qubit and for nonideal implementation of logic gates \cite{bau05,rya08,par15,ata16,zai21}, for arbitrary cycles numbers including the transient regime. We obtain explicit expressions for the fundamental upper bounds for COP and cooling power in the ideal reversible limit and compare the former to the ideal Carnot COP of a quantum refrigerator \cite{kos14,rez09,all10,aba16}. We finally experimentally determine the performance of the minimal algorithmic cooling refrigerator using three qubits in a nitrogen-vacancy  (NV) center in diamond \cite{zai21} and obtain values of COP and cooling power that are close to their fundamental bounds.

\paragraph{Quantum algorithmic cooling refrigerator.} We consider a minimal three-qubit heat-bath algorithmic cooling refrigerator with Hamiltonian \(H= \sum_i \gap{i}\), where \(\omega_i\) is the frequency and \(\sigma^z_i\) the usual Pauli operator of each spin. Qubit 1 is the target spin while qubits 2 and 3 are the two reset spins.  The machine starts in a separable state of the  three qubits, \(\rho(0) = \otimes_i \rho_i(0)\), with  respective density matrices \(\rho_i (0)= \text{diag}(1-\epsilon_i(0), 1+ \epsilon_i(0))/2\) and  polarizations \(\epsilon_i(0)\). {We  denote by   \(\tilde \rho_i(n)\) the various states after \(n\) iterations of the compression stage  and  by \(\rho_i(n)\) those after both compression and refresh steps.} We next identify the  heat \(Q(n)\) extracted during round \(n\) with the  average energy change of the target qubit, {\(Q(n) = \tr\{ \gap{1} [\rho_1(n+1) - \rho_1( n) ] \}\)}. We further associate the work performed by the logic gates on the system with the corresponding mean energy variation,  {\(W(n)= \sum_i \tr\{ \gap{i} [ \tilde\rho_i(n+1) - \rho_i( n ) ] \}\)} \cite{rem07}. The COP per cycle, \(\zeta(n)\), is then defined as the ratio of  pumped heat and  applied work, while the cooling power per cycle, \(J(n)\), is given (in units of  the cycle time) as the discrete derivative (or {forward} difference) of the heat:
\begin{equation}
\label{1}
\zeta(n) = -\frac{Q(n)}{W(n)} \quad \text{and} \quad {J(n) = Q(n+1)-Q(n)}.
\end{equation}
These are the principal quantities of our investigation.

We shall  examine the  thermodynamic properties of heat-bath algorithmic cooling in the  general  case  where compression is implemented with imperfect gates and the (slow-relaxing) target spin is subjected to irreversible energy dissipation  \cite{com}. We will discard   irreversible losses of  the reset spins because of their much faster relaxation.
For each round \(n\) of the cooling protocol, we accordingly describe the evolution of the system with the help of three quantum channels  \cite{nie00}. We first account for  energy  dissipation of the target qubit via an amplitude damping channel \(\mathcal{D}\) with decay rate \(\gamma\) \cite{nie00},
 \begin{equation}
{\mathcal{D}[\bullet] =\sum_{j = 1, 2}
    \Gamma_j \bullet \Gamma_j^\dag, }
\end{equation}
with  the two Kraus damping operators,
\begin{equation} \label{eq:damp-kraus}
    \Gamma_1 = \begin{pmatrix}
        1 & 0 \\ 0 & \sqrt{1 - \gamma}
    \end{pmatrix}  \quad \text{and} \quad
    \Gamma_2 = \begin{pmatrix}
        0 & \sqrt{\gamma} \\ 0 & 0
    \end{pmatrix}.
\end{equation}
We further characterize the imperfect compression stage with the  channel {\(\tilde \rho(n)= \mathcal{C}[\rho(n-1)] \)}, such that,
\begin{equation} \mathcal{C}[\bullet] =
    \sum_{k = 1, 2} K_k  \bullet  K_k^\dag,
    \end{equation}
where we have introduced the two quantum operators,
\begin{eqnarray}
K_1 =&& \frac{I}{\sqrt{2}} - \frac{1}{\sqrt{2}} ( \ketbra{011}{011} + \ketbra{100}{100} )\nonumber \\
&&- \mathrm{i}\, ( \sin\theta \ketbra{011}{100} + \text{h.c.} ), \\
K_2 =&& \frac{I}{\sqrt{2}} + \left( \cos\theta - \frac{1}{\sqrt{2}} \right) \ketbra{011}{011}\nonumber \\
&& -\left( \cos\theta - \frac{1}{\sqrt{2}} \right) \ketbra{100}{100} .
\end{eqnarray}
Here \(\ket{0}\) and \(\ket{1}\) are the eigenstates of the spin operators \(\sigma^z_i\) and \(I\) denotes the unit operator. The angle \(\theta\) quantifies the imperfection of the compression step. When \(\theta= \pi/2\), we recover  ideal compression which swaps the diagonal elements of the density matrix, \(U= \exp(-\mathrm{i} \pi  V/2)\) with \(V=\ketbra{100}{011} + \ketbra{011}{100}\) \cite{boy02,par16,fer04,sch05,sch07,rem07,kay07,bra14,rai15,rod16,rai19,rai21}. The compression operation is commonly  implemented experimentally with Toffoli or CNOT gates with imperfect fidelity, which leads \(\theta\) to deviate from the  ideal value \(\pi/2\) \cite{bau05,rya08,par15,ata16,zai21}. We finally describe the refresh step through \cite{boy02,par16,fer04,sch05,sch07,rem07,kay07,bra14,rai15,rod16,rai19,rai21},
\begin{equation}
{\rho(n) =\mathcal{R}[ \tilde \rho(n) ] = \tr_{23}\{ \tilde \rho(n) \} \otimes \rho_2(0) \otimes \rho_3(0).}
\end{equation}
 The composition of the above three channels yields the combined {quantum operation \(\mathcal{E}[\bullet]\)} which corresponds to one round of the refrigeration algorithm.

 \begin{figure}[t]
	\centering
	\includegraphics[width=0.40\textwidth]{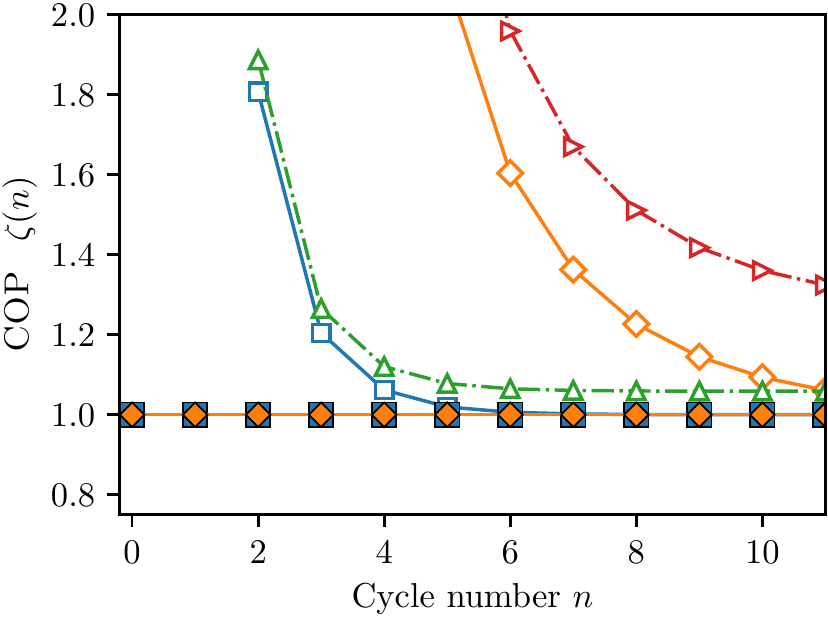}
	\caption{Comparison with the Carnot coefficient of performance. In the  reversible regime (\(\gamma=0\)), the  coefficient of performance \(\zeta(n)\) (full symbols) gets close to the corresponding Carnot limit \(\zeta_\text{C}(n)\) (empty symbols) after a few cycles. The Carnot bound is  generally not reached in the presence of  losses (\(\gamma \neq 0\)). {Same parameters as in Fig.~2.}}\label{fig:appfig2}
\end{figure}
 \paragraph{Analytical results.} We analytically solve the  dynamics generated by the quantum channel \(\mathcal{E}[\bullet]\) for an arbitrary number \(n\) of algorithmic cooling cycles, using vectorization techniques in Liouville space \cite{gya20}. {In this approach, a density matrix \(\sigma\) is mapped onto a vector} \(\opn{vec}(\sigma)\) (often called supervector) in a higher-dimensional Hilbert space, \( \sigma = \sum_{r, s} \sigma_{rs} \ketbra{r}{r} \mapsto \opn{vec}({\sigma}) = \sum_{r, s} \sigma_{rs} \ket{r}\ket{s}\), where the index \(r\) is varied  first. The  quantum channel, with   operator-sum representation \(\mathcal{E}[\sigma] = \sum_\mu E_\mu \sigma E_\mu^\dag\), may then be expressed as  \(\mathcal{E}(\sigma) = \opn{unvec}\big(\Phi_\mathcal{E} \opn{vec}(\sigma)\big)\) with the superoperator \(\Phi_\mathcal{E}=\sum_\mu E_\mu \otimes (E_\mu^\dag)^\intercal\). The advantage of the Liouville space representation is that \(n\) iterations of the cooling cycle may be simply evaluated by computing \(\Phi_\mathcal{E}^n\), which is not possible in the original Hilbert space (see Supplemental Material \cite{sup}). Using this formalism, we obtain explicit expressions for the polarization of the target qubit, as well as for heat and work, from which we deduce COP and cooling power \eqref{1} for each cycle, for arbitrary initial {polarizations of the three qubits} \cite{sup}.

 For simplicity, we here indicate the formulas for  COP and  cooling power for the experimentally relevant case of vanishing initial polarization of the target qubit  \cite{bau05,rya08,par15,ata16,zai21}:
 \begin{widetext}
 \begin{eqnarray}
    \zeta(n) ={}&& \frac{
        -[ 2\gamma ( 1 + \cos^2\theta ) - 2\epsilon  (2 + \gamma \epsilon) \sin^2\theta ] [ (\gamma - 1) f(\theta) + 4 ]
    e^{-ng(\theta, \gamma) }}{
        [ (\gamma - 1) ( f(\theta) + 4 (\epsilon^2 + 1) \sin^2\theta ) + 4]
        [ \gamma f(\theta) + 2\epsilon (\cos(2\theta) - 1) ] e^{-ng(\theta, \gamma) }
        + 16 (1 + \epsilon)^2 \gamma \sin^2\theta
    }
     \xrightarrow[]{\gamma=0} \zeta_\text{max}(n) = 1 \nonumber \\
    J(n) ={}&& {\frac{1}{16} [ (\gamma - 1) f(\theta) + 4 ] [ 4\epsilon \sin^2\theta - \gamma f(\theta) ] e^{- ng(\theta, \gamma)} \xrightarrow[\theta=\theta_n]{\gamma=0} J_\text{max}(n)= \frac{\epsilon}{2} (1 +\epsilon^2)e^{- n g(\theta_n, 0)}},
    \end{eqnarray}
\end{widetext}
{where we have defined the two functions \( f(\theta) = 3 + (1 + \epsilon^2) \cos(2\theta) - \epsilon^2 \) and
\(g(\theta, \gamma) = \ln[ {4}/{( (1 - \gamma) f(\theta)) } ]\), and   introduced  the angle  \(\theta_n= \pi/2\) for \(n <2\), \(\epsilon < \sqrt{1/3}\) and \(\theta_n =  \arccos[(2 \epsilon^2 + n \epsilon^2 + n - 6)/((2 + n) (1 + \epsilon^2)) ]/2\) otherwise. We have here set  \(\epsilon_1(0)=0\),
\(\epsilon_2(0) = \epsilon_3(0) = \epsilon\)
 (results for general polarizations are given in Ref.~\cite{sup})}.

Figures 2ab) represent \(\zeta(n)\) and \(J(n)\) as a function of the cycle number \(n\) for various values of the decay rate \(\gamma\) and of the  mixing angle \(\theta\). We first note that both quantities reach their fundamental maximum values in the undamped limit \(\gamma=0\). In this unitary, reversible regime, the COP \(\zeta(n)\) is equal to  one, implying that the extracted heat is precisely given by the work supplied by  the gate operations, \(-Q(n)= W(n)\) (when \(\gamma=0\)). The value of \(\zeta_\text{max}(n)\) is  moreover independent of the  cycle number \(n\) and of the  angle \(\theta\). This interesting point  reveals that gate imperfections do not affect the maximum efficiency of the algorithmic cooling refrigerator, but only reduce   the power \(J_\text{max}(n)\). We further observe that the cooling power generically decays exponentially to zero with increasing cycle iterations, {as the asymptotic temperature is reached and no more heat can be extracted from the target qubit}---a behavior also exhibited by \(\zeta(n)\) in the presence of irreversible losses. Figure 2b additionally shows that  \(J(n)\) is mostly affected by  the angle \(\theta\) and not so much by  the decay rate \(\gamma\) {in the experimentally relevant range \(\gamma < 0.01\)}. {In particular, the optimal angle \(\theta_n\) in \(J_\text{max}(n)\) depends on \(n\) for \(n \geq 2\) \cite{com2}.}

Two important features of the algorithmic cooling protocol are the asymptotic polarization of the target qubit and the number of iterations needed to reach it \cite{boy02,par16,fer04,sch05,sch07,rem07,kay07,bra14,rai15,rod16,rai19,rai21}. Using the Liouville space solution, we find the exact expression (again for \(\epsilon_1(0)=0\), \(\epsilon_2(0) = \epsilon_3(0) = \epsilon\)) \cite{sup},
\begin{eqnarray}
\epsilon_1(n) =&&
\frac{
    \gamma f(\theta) + 2\epsilon [\cos(2\theta) - 1]
}{
    (\gamma - 1)f(\theta) + 4
} [ e^{-ng(\theta,\gamma)} - 1 ]\nonumber \\
&&\xrightarrow[\theta=\pi/2]{\gamma=0} {{\epsilon_1}_\text{max}(n)}=\frac{2\epsilon}{1+\epsilon^2} [1- e^{-ng(\pi/2,0)}  ].
\end{eqnarray}
The stationary value \(\epsilon_1(\infty)\) is thus approached exponentially  with a rate constant  given by \(1/g(\theta,\gamma)\). Figure 2c) displays a radically different effect of energy dissipation and of gate imperfection on the nonideal polarization of the target qubit. While the decay constant \(\gamma\) affects the asymptotic value of the polarization \(\epsilon_1(\infty)\), the mixing angle \(\theta\) modifies the convergence rate to that value {for \(\gamma=0\)}. As a consequence, imperfect gate operation does not prevent achieving maximum polarization in the reversible limit, it only increases the number of required iterations. {This property holds for all convex combinations of the ideal compression and the identity \cite{sup}.}

\begin{figure}[t]
\centering
\begin{tikzpicture}
\node (a) [label={[label distance=-.7 cm]145: \textbf{a)}}] at (-0.3,0) {\includegraphics[width=0.415\textwidth]{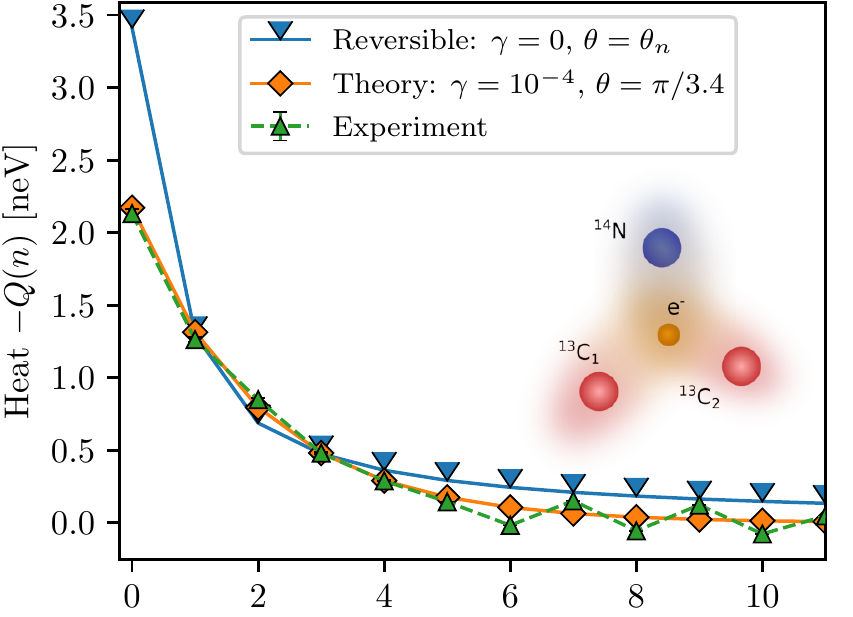}};	\node (a) [label={[label distance=-.7 cm]145: \textbf{b)}}] at (-0.37,-5.6) {\includegraphics[width=0.411\textwidth]{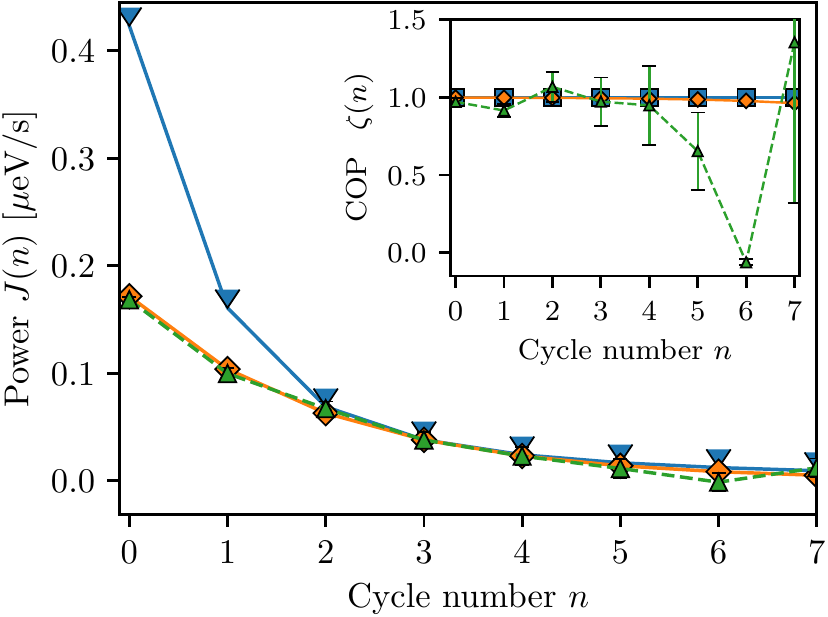}};
\end{tikzpicture}
\caption{Experimental performance of the three-qubit algorithmic cooling refrigerator. a) Experimental data for  heat \(Q(n)\)  (green triangles) show excellent agreement with theory (orange diamond) with \(\gamma = 10^{-4}\) and \(\theta = \pi/3.4\). b) Cooling power \(J(n)\) and COP \(\zeta(n)\) also  agree very well with theory {(\(\zeta(n)\) becomes sensitive to measurement errors for larger \(n\))}. {Error bars correspond to the standard deviation.}}
\end{figure}

Let us next compare the thermodynamic performance of the algorithmic cooling refrigerator to that of a conventional quantum refrigerator \cite{kos14,rez09,all10,aba16}, whose COP  is upper bounded by the Carnot formula, \(\zeta_\text{C} = T_\text{c}/(T_\text{h}- T_\text{c})\). We accordingly evaluate, for each cycle \(n\), the temperature of the target qubit via {\( T_\text{c}(n) = 1 / \ln[ \boldsymbol{(} 1 + \epsilon_1(n) \boldsymbol{)} / \boldsymbol{(} 1 - \epsilon_1(n) \boldsymbol{)} ] \)}, {determined via the ratio of the (Boltzmann distributed) populations of excited and ground states}  (a similar formula holds for the initial hot temperature of the reset spins). The corresponding Carnot COP per cycle, \(\zeta_\text{C}(n) = T_\text{c}(n)/[T_\text{h}- T_\text{c}(n)]\) for the algorithmic cooling refrigerator is shown, together with the COP \(\zeta(n)\), Eq.~(8), {in Fig.~3}). While \(\zeta(n)\) is smaller than \(\zeta_\text{C}(n)\) at the beginning of the refrigeration cycle, the Carnot bound is quickly approached after only a few iterations in the ideal limit \((\gamma=0,\theta=\pi/2)\). The Carnot limit is in general not attained in the presence of damping (\(\gamma \neq 0\)).

\paragraph{Experimental results.} {We finally experimentally validate our new theoretical framework, and analyze the performance of an algorithmic cooling refrigerator made of  three nuclear spins that are hyperfine coupled to the central electron spin of a  NV center in diamond \cite{zai21}}. NV center systems  offer excellent control of their states and exhibit very long spin coherence times \cite{doh13}. The target spin and the two  reset spins are respectively chosen to be the Nitrogen \(^{14}\)N and two Carbon \(^{13}\)C nuclear spins that are coupled to the central electron spin of the NV center with respective strengths \(2.16\)MHz, \(90\)kHz, and \(414\)kHz (Fig.~4). The central electron spin has a twofold role: it acts as (i) the heat bath and also as (ii)  an ancillary spin that drives the interaction among the spins required to achieve the entropy compression on the target qubit \cite{zai21}. The optical spin polarization of the central NV-spin is transferred to the two \(^{13}\)C spins via a SWAP gate during the refresh steps \cite{sup}.
Compression is implemented by performing a non-local gate among the three nuclear spins that allow for a unitary mixing of populations in the subspace of \([\ket{011}, \ket{100}]\) \cite{sup}. As the nuclear spins do not interact with each other, this three qubit Toffoli gate is mediated by the electron spin.

Typical times of  each  step  are   \({\sim}285\mu\)s for the compression step and   \({\sim}5\)ms for the refresh step.
The life-time  of the  nuclear spin, \(T_1\),  is of the order of seconds (corresponding to a decay rate \(\gamma\simeq 10^{-4}\)),  allowing us to perform multiple rounds of the cooling cycle. Since the refresh step periodically resets the two \(^{13}\)C spins, their damping is not relevant over the duration of the experiment. Another source of noise,  not considered in previous experiments \cite{bau05,rya08,par15,ata16},  is  due to imperfect compression: the compression algorithm indeed utilizes  three-qubit Toffoli gates \cite{zai21}, which when transpiled into the electron-nuclear spin gates, involve \(5\) CNOT gates and \(14\) single-qubit rotations. Gate imperfections, together with  imperfect charge state initialization, lead to mixing between the   states \(\ket{011}\) and \(\ket{100}\), which can be quantified by an effective mixing angle \(\theta\). The best fit in our experiment is \(\theta\simeq \pi/3.4\), which corresponds to an overall error of \( {\sim}20\%\) in the compression step. Reset is additionally  implemented via an iterative SWAP gate that allows for a \({\sim}99\%\) fidelity on the achievable hot spin polarization.

 The  initial polarizations of the  two reset spins  are {\(\epsilon_2(0) {\sim} 0.58\) and \(\epsilon_3(0){\sim} 0.41\)}. The imbalance between the  polarizations  comes from the different coupling strengths of the two spins to the electron spin. We measure the target spin polarization via single-shot readout with a fidelity of \({\sim}97\%\), from we which we evaluate heat \(Q(n)\) and cooling power \(J(n)\), {as well as work \(W(n)\) and   COP \(\zeta(n)\)} for each cycle  (see Supplemental Material \cite{sup}) \cite{coml}. We obtain excellent agreement between theory (with \(\gamma= 10^{-4}\) and \(\theta=\pi/3.4\)) and data (Figs.~4ab).
 {We observe especially  that the upper bounds \(J_\text{max}(n)\) and \(\zeta_\text{max}(n)\), given by Eq.~(8),  are reached in the experiment. For \(n\geq 5\), heat and work are very small. As a result, the COP becomes highly  sensitive to measurement errors: it can get negative for \(-Q(n)\) below zero (as for \(n= 6\)) or be larger than one if \(W(n)\) is too close to zero (as for \(n= 7\)).}

 \paragraph{Conclusions.} We have performed a theoretical and experimental study of the thermodynamic performance of a minimal three-qubit algorithmic cooling refrigerator. We have analytically computed key figures of merit, such as coefficient of performance, cooling power and polarization of the target qubit, for arbitrary cycle number. We have determined their fundamental upper bounds in the ideal reversible limit and shown that   the coefficient of performance quickly convergences to the  Carnot value after a few cycles. We have further highlighted the effects of realistic experimental imperfections, in particular, irreversible energy dissipation of the target qubit and imperfect gate operations, on these quantities. We have finally demonstrated that the fundamental limits may be approached in an experimental system made of the three qubits of a NV center in diamond.

 \begin{acknowledgments}
    We acknowledge financial support by the DFG (FOR 2724), European Union's Horizon 2020 research and innovation program ASTERIQS under grant No. 820394, European Research Council advanced grant No. 742610, SMel, Federal Ministry of Education and Research (BMBF) project MiLiQuant and Quamapolis, the Max Planck Society, and the Volkswagentiftung. R. R. Soldati acknowledges the financial support by the German Academic Exchange Service (DAAD) research grant Bi-nationally Supervised Doctoral Degrees/Cotutelle, 2020/21 (57507869). We would like to acknowledge J. Meinel and S. Zaiser for their support on graphical representations.
 \end{acknowledgments}

\pagebreak
\clearpage
\widetext

\appendix

\section{Supplemental Material: Thermodynamics of  a minimal algorithmic cooling refrigerator}

The Supplemental Material contains details about (I) the  analytical solution of the dynamics of the three-qubit heat bath algorithmic cooling refrigerator for arbitrary cycle number, (II) the theoretical evaluation of heat, work, target qubit polarization, as well as cooling power and coefficient of performance  for arbitrary initial polarizations, {(III) the general properties of the imperfect compression map, (IV) the experimental implementation of the cooling algorithm, and (V) the experimental evaluation of the thermodynamic performance of the refrigerator.}

\section{Analytical solution of the qubit dynamics in Liouville space}
This section presents the exact solution of the dynamics of the three-qubit system in Liouville space \cite{gya20}. We begin with a brief reminder of the   vectorization technique in order to set the notation \cite{gil09}.

\subsection{Vectorization method}
The Liouville representation provides a compact and efficient method to describe quantum processes by mapping operators in Hilbert space onto vectors in an enlarged space.
Let the space of operators on a Hilbert space be \(\mathsf{B}(\mathbf{H})\). Let us further consider its representation in terms of the Hilbert space and its dual, \(\mathsf{B}(\mathbf{H}) \cong \mathbf{H} \otimes \mathbf{H}^*\), or in other words, in terms of bras and kets. Vectorization is then the isomorphism mapping
\begin{equation}
    \opn{vec} \colon \mathbf{H} \otimes \mathbf{H}^*
    \to
    \mathbf{H} \otimes \mathbf{H} ,
\end{equation}
where we define the Liouville space as \(\mathbf{L} = \mathbf{H} \otimes \mathbf{H}\).
This map naturally extends to linear maps on the space of operators itself. In the case of linear transformations, such as the operator-sum representation of quantum channels, we have \(\mathcal{M}[\bullet] = \sum_\mu M_\mu \bullet M_\mu^\dag \in \mathsf{B}(\mathbf{H}) \otimes \mathsf{B}(\mathbf{H}^*)\), with \(\mathsf{B}(\mathbf{H}^*) \cong \mathsf{B}(\mathbf{H})^*\). Adopting the character of Hilbert space of the spaces of operators themselves, vectorization extends to them yielding
\begin{equation}
    \text{(extended)}\quad
    \opn{vec} \colon \mathsf{B}(\mathbf{H}) \otimes \mathsf{B}(\mathbf{H}^*)
    \to
    \mathsf{B}(\mathbf{H}) \otimes \mathsf{B}(\mathbf{H}) .
\end{equation}
Since we are working with qubit spaces, we have  \(\mathbf{H} = \mathbb{C}^2\) and \(\mathsf{B}(\mathbf{H}) = \mathbb{M}^{2 \times 2}_\mathbb{C}\), where \(\mathbb{M}^{2 \times 2}_\mathbb{C}\) is the space of 2-by-2 matrices with complex entries. Starting from these building blocks, vectorization is the same as rearranging columns and rows of matrices, with different ways of doing this related by reshuffling their components.

With the conventions taken in the main text, the action of vectorization on density matrices is to stack their columns in a single-column vector  \cite{hor12}. We denote the vectorized density matrices as \(\vec{\rho} = \opn{vec}(\rho)\). Quantum channels, in operator-sum representation with Kraus operators \(E_\mu\), are accordingly mapped to the matrices
\begin{equation} \label{eq:supermatrix}
    \Phi_\mathcal{M} = \sum_\mu M_\mu \otimes (M_\mu^\dag)^\intercal ,
\end{equation}
that act on \(\vec\rho\) from the left as regular matrix multiplication. These rules extend to the states and channels on the tensor product of target and two reset qubits Hilbert spaces, the composite Liouville state is thus \(\mathbf{L} = \mathbf{L}_1 \otimes \mathbf{L}_2 \otimes \mathbf{L}_3\).

\subsection{Solutions for target and reset qubits}

In order to evaluate the states  of the target qubit and of the two reset qubits after an arbitrary number of refrigeration cycles \(n\), we need to compute the matrix in Liouville space of the combined quantum channel consisting of damping \(\mathcal{D}\), compression \(\mathcal{C}\) and refresh \( \mathcal{R}\) maps.  As a first step, we treat the damping channel \(\mathcal{D}[\bullet]= \sum_{j} \Gamma_j \bullet \Gamma_j^\dag \), given by Eqs.~(2)-(3) of the main text, and use formula \eqref{eq:supermatrix}  to build the corresponding Liouville superoperator \(\Phi_\mathcal{D} = \sum_j \Gamma_\mu \otimes (\Gamma_\mu^\dag)^\intercal\), a 4-by-4 matrix acting  on the target subspace.

As a second step, we consider the compression channel, given by Eqs.~(4)-(6) of the main text. {Since we focus on the target qubit for the time being, we introduce the  reduced compression channel  \(\mathcal{C}_\text{red}[\rho_1] = \tr_{23}\big\{ \sum_{k = 1, 2} K_k \big( \rho_1 \otimes \rho_2(0) \otimes \rho_3(0) \big) K_k^\dag \big\}\), that acts on the target qubit space alone.}
  {We determine the Kraus operators \(C_j\) of the  operator-sum representation $\mathcal{C}_\text{red}[\bullet]= \sum_{j}
    C_j \bullet  C_j^\dag$} by extending the treatment  that can be found in Ref.~\cite{nie00}, Section 8.2.3, to nonunitary operations. It involves introducing a purification of the mixed state of reset qubits, that is, the map
\begin{equation}
    \rho_2(0) \otimes \rho_3(0) \mapsto \ket{\rho_2, \rho_3} \in \mathbf{H}_2 \otimes \mathbf{H}_3 \otimes \mathbf{H}_\text{R} ,\quad \text{with }
    \ket{\rho_2, \rho_3} = \sum_{ij} \ket{ij} \ket{ij}_\text{R} \sqrt{ \bra{i}\rho_2(0)\ket{i} \bra{j}\rho_3(0)\ket{j} } ,
\end{equation}
such that \(\tr_\text{R} \ketbra{\rho_2, \rho_3}{\rho_2, \rho_3} = \rho_2(0) \otimes \rho_3(0)\), where \(\mathbf{H}_\text{R}\) is an artificial, reference Hilbert space introduced as part of the purification. By explicitly evaluating  the trace and the purification, we find,
\begin{equation}
    C_{(k), ij}^{i'j'} = \bra{i'j'} \bra{ij}_\text{R} \big( K_k \otimes I_\text{R} \big) \sum_{rs} \sqrt{ p_2^r p_3^s } \ket{rs}\ket{rs}_\text{R} ,\quad \text{with } p_2^r = \bra{r}\rho_2(0)\ket{r},\ p_3^s = \bra{s}\rho_3(0)\ket{s} ,
\end{equation}
where \(I_\text{R}\) is the identity on the reference Hilbert space.  The bras here stem from the trace while  the sum over double-primed indices and their kets stem from the purified state. This leads to the operators%
\begin{equation}
    C_{(k), ij}^{i'j'} = \bra{i'j'} K_k \ket{ij} \sqrt{p_2^i p_3^j} .
\end{equation}
To simplify the notation, we group the indices \(i, j, i', j'\) and \(k\), each binary, into the new index \(\mu\) ranging over 32 values, and thus yielding 32 operators. Most of them are identically 0 through this procedure, however, with six others remaining for \(k = 1\) and other four for \(k = 2\). The corresponding superoperator  \(\Phi_\mathcal{\tilde C}\) is again given by Eq.~\eqref{eq:supermatrix}.

\begin{figure}[t]
    \centering
    \includegraphics{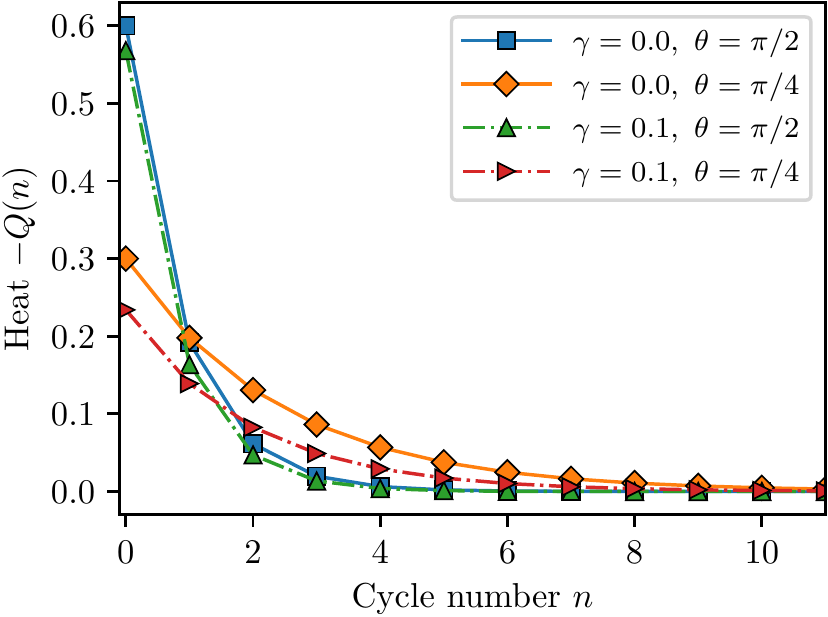}
    \includegraphics{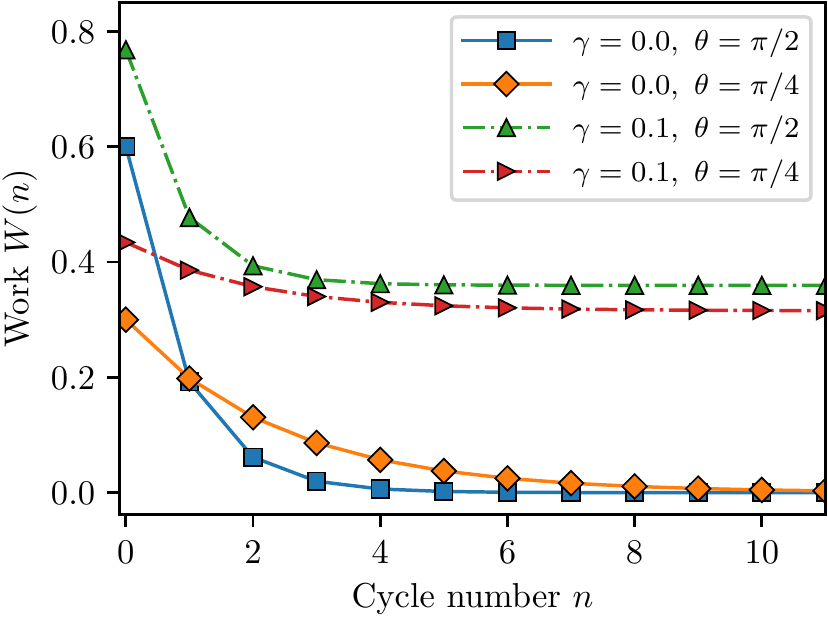}
    \caption{Heat \(Q(n)\), Eq.~\eqref{12}, and work \(W(n)\), Eq.~\eqref{16}, as a function of the number \(n\) of cycles, for various values of damping coefficient \(\gamma\) and of the mixing angle \(\theta\). Parameters are \(\epsilon_1(0) = 0\), \(\epsilon_2(0)=\epsilon_3(0)=0.6\).}
    \label{fig:qw}
\end{figure}
The  state of the target qubit, which is needed to determine heat and cooling power of the heat bath algorithmic cooling refrigerator, is obtained after  concatenation of the two maps \(\mathcal{D}\) and \(\mathcal{C}_\text{red}\), which in vectorized form reads  \(\Phi_{\mathcal{C}_\text{red}} \Phi_\mathcal{D}\). The target qubit states after a number \(n\) of cycles are then obtained by calculating \((\Phi_{\mathcal{C}_\text{red}}\Phi_\mathcal{D})^n\). We find  for arbitrary initial polarizations \(\epsilon_1(0)\), \(\epsilon_2(0)\) and \(\epsilon_3(0)\)
\begin{equation}
\label{9}
   \tilde \rho_1(n)
    = \opn{unvec}\big\{ \big( \Phi_{\mathcal{C}_\text{red}} \Phi_\mathcal{D} \big)^n \vec\rho_1(0) \big\}
    = \frac{1}{2}
    \begin{pmatrix}
        1 - \epsilon{1}(n) & 0 \\ 0 & 1 + \epsilon{1}(n)
    \end{pmatrix}.
\end{equation}
The polarization \(\epsilon{1}(n, \theta, \gamma)\) is here explicitly given by
\begin{equation} \label{eq:target-qubit}
    \epsilon{1}(n) =
    \frac{
        2 (\epsilon{2}(0) + \epsilon{3}(0)) \sin^2\theta - \gamma F(\theta)
        + \big[ 2 \sin^2\theta \big( (1 + \epsilon{2}(0) \epsilon{3}(0)) \epsilon{1}(0) - \epsilon{2}(0) - \epsilon{3}(0) \big) + \gamma (1 + \epsilon{1}(0)) F(\theta) \big] e^{-nG(\theta,\gamma) }
    }{ ( \gamma - 1 ) F(\theta) + 4 }
\end{equation}
where
\begin{equation}
\begin{split}
    F(\theta) ={}& 3 + (1 + \epsilon{2}(0)\epsilon{3}(0)) \cos(2\theta) - \epsilon{2}(0)\epsilon{3}(0) \xrightarrow{\epsilon{2}(0) = \epsilon{3}(0)} f(\theta) , \\
    G(\theta, \gamma) ={}& \ln\left( \frac{4}{ (1 - \gamma) F(\theta) } \right) \xrightarrow{\epsilon{2}(0) = \epsilon{3}(0)} g(\theta, \gamma) .
\end{split}
\end{equation}
These expressions are symmetric under exchange of the initial polarizations of the two reset qubits (\(\epsilon{2}(0) \leftrightarrow \epsilon{3}(0)\)). {The ideal asymptotic polarization \(2\epsilon/(1+\epsilon^2)\), obtained for \(\gamma=0\), \(\theta=\pi/2\) and  \(\epsilon_1(0)=0\), \(\epsilon_2(0) = \epsilon_3(0) = \epsilon\), agrees with the one derived in Refs.~\cite{rod16,zai21}. For different reset spin polarizations, this ideal limit gets modified to \([\epsilon_2(0)+ \epsilon_3(0)]/[1+\epsilon_2(0) \epsilon_3(0)]\). }

The quantum map \(\Phi_{\mathcal{C}_\text{red}} \Phi_\mathcal{D}\) is found without difficulties because the compression step is evaluated  with respect to a tensor product with fixed reset states, allowing for the method of Ref.~\cite{nie00} to be applied. However, this simplification does not occur when changing reset qubits are involved, as in the refresh operation.

\begin{figure}[t]
    \centering
    \includegraphics{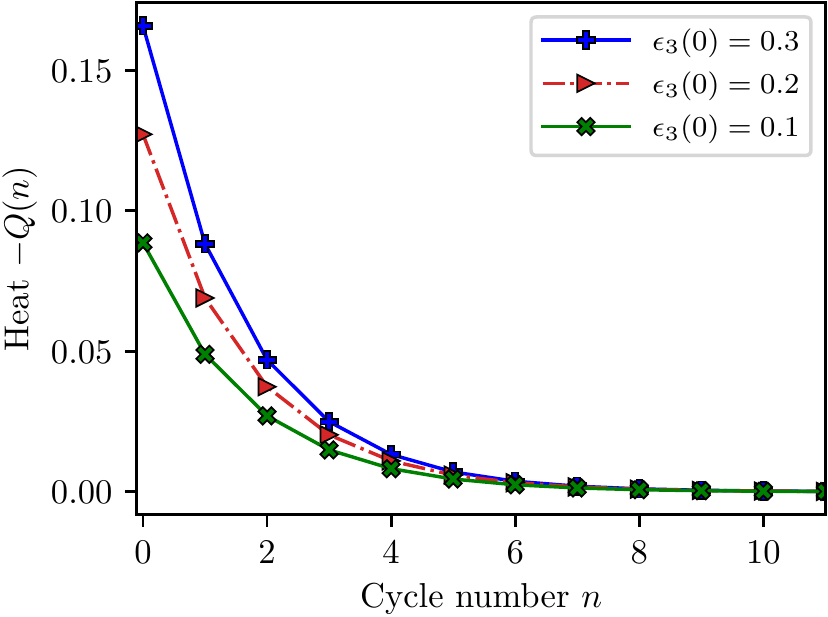}
    \includegraphics{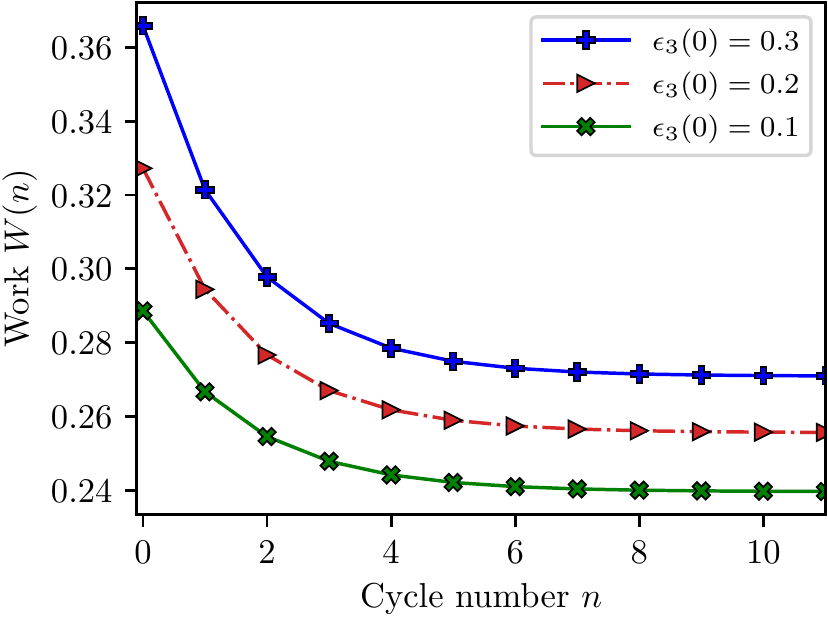}
    \includegraphics{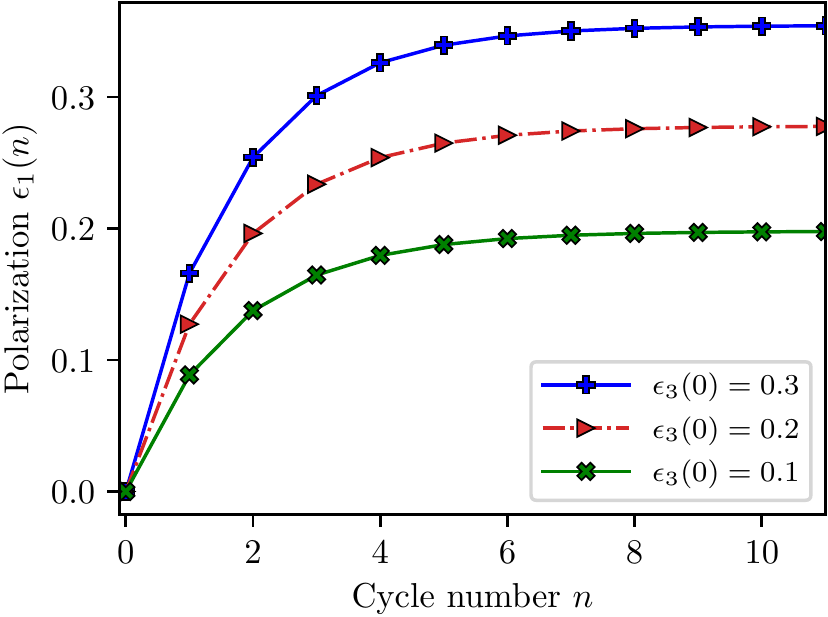}
    \includegraphics{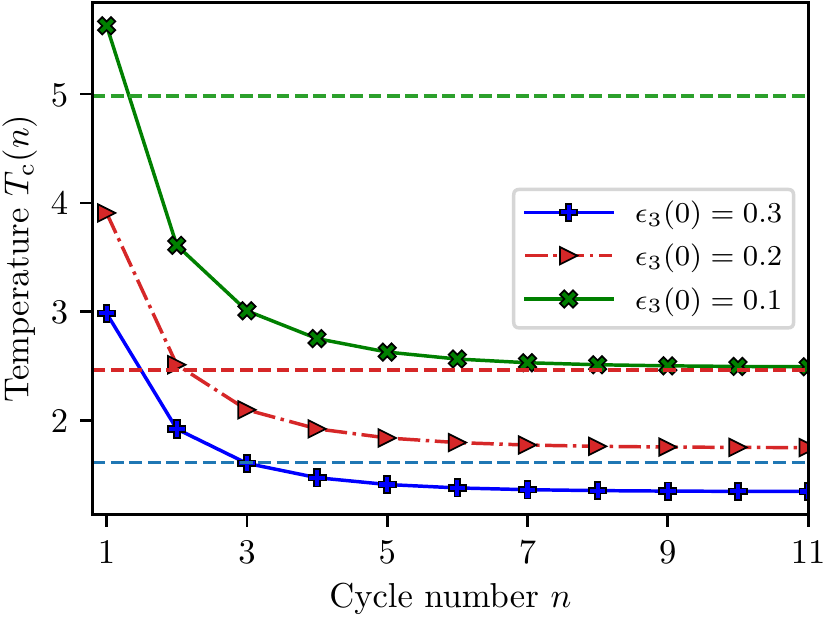}
    \caption{Heat \(Q(n)\), Eq.~\eqref{12},  work \(W(n)\), Eq.~\eqref{16}, and polarization of the target qubit \(\epsilon_1(n)\), Eq.~\eqref{9}, as well as the corresponding temperature \(T_c(n)\),  as a function of the number \(n\) of cycles, for fixed values of the damping coefficient \(\gamma=0.1\) and of the mixing angle \(\theta=\pi/3\), for different initial polarizations of the reset qubit 3. Parameters are \(\epsilon_1(0) = 0\), \(\epsilon_2(0) =0.3\). Changing the initial polarization of the reset qubit may either increase or decrease the values of these thermodynamic quantities.
        Dashed lines in the lower right plot correspond to the respective temperatures of the reset qubits with matching colors (the red line also corresponds to the polarization \(\epsilon_2(0)\)).
    }
    \label{fig:diffpol}
\end{figure}

As a next step, we deal with the refresh operator \(\mathcal{R}[ \bullet ]\) given by Eq.~(7) of the main text. To that end, we consider an extension \(\mathcal{D}_\text{ext}\) of the damping channel that acts trivially (through the identity) on the reset qubits \(2\) and \(3\). We accordingly define the channel \(\mathcal{E}=\mathcal{R}[ \mathcal{C}[ \mathcal{D}_\text{ext}[\bullet] ] ]\)
 for the ensemble of three qubits, as
\begin{equation} \label{eq:full-state}
    \rho(n) = \mathcal{E}[ \rho(n-1) ]
    = \mathcal{R}\left[ \sum_k K_k \big( \mathcal{D}_\text{ext} [\rho(n-1)] \big) K_k^\dag \right],
\end{equation}
where \(\rho(n) = \rho_1(n) \otimes \rho_2(0) \otimes \rho_3(0)\). The role of the refresh operation is to keep the reset states equal to \(\rho_2(0) \otimes \rho_3(0)\) at the beginning and at the end of each cooling cycle. We denote the corresponding superoperator as \(\Phi_\mathcal{E}\).

The determination of the  coefficient of performance and, in turn, of the work applied during the compression stage requires the knowledge of the reset qubit states before the refresh stage. The latter are given by
\begin{equation} \label{eq:bath-qubit}
\begin{split}
    \tilde \rho_i(n) = \tr_{1, j \neq i}\big\{
    \sum_k K_k \big( \mathcal{D}_\text{ext} [\rho(n-1)] \big) K_k^\dag\big\}
    = \frac{1}{2}
    \begin{pmatrix}
        1 - \epsilon{i}(n) & 0 \\
        0 & 1 + \epsilon{i}(n)
    \end{pmatrix} ,
\end{split}
\end{equation}
where \(i\) and \(j\) are either 2 or 3, {and \(\rho(n-1) = \rho_1(n-1) \otimes \rho_2(0) \otimes \rho_3(0)\)}. The polarizations \(\epsilon{i}(n)\) explicitly read
\begin{multline}
    \epsilon_i(n) =
    \frac{
        2 I_i(\gamma) \sin^2\theta + 4\gamma ( 1 - \epsilon{i}(0) \cos^2\theta )
    }
    { 2 [ ( \gamma - 1 ) F(\theta) + 4 ] } \\
    + \frac{
        \sin^2\theta ( 1 + \epsilon{2}(0) \epsilon{3}(0) ) }{F(\theta)} \frac{
        \big[ 2 \sin^2\theta \big( (1 + \epsilon{2}(0) \epsilon{3}(0)) \epsilon{1}(0) - \epsilon{2}(0) - \epsilon{3}(0) \big) + \gamma (1 + \epsilon{1}(0) F(\theta) \big]
    }
    { ( \gamma - 1 ) F(\theta) + 4 } e^{-nG(\theta,\gamma)} ,
\end{multline}
with
\begin{equation}
    I_i(\gamma) = \big( (\gamma - 1) \epsilon{i}(0)^2 + \epsilon{i}(0) + \gamma \big) \epsilon{j \neq i}(0) - \epsilon{i}(0) + 1 .
\end{equation}
{The reset  qubit states only change within a single stroke (before they are refreshed). Their dependence upon the \(n-1\) previous applications of the cooling cycle is implicit in the target qubit input state \(\rho_1(n-1)\). This is schematically represented as \(\epsilon{1}(n) \to \epsilon{1}(n+1)\) while \(\epsilon{2, 3}(0) \to \epsilon{2, 3}(n+1)\).}

\begin{figure}[t]
    \centering
    \includegraphics{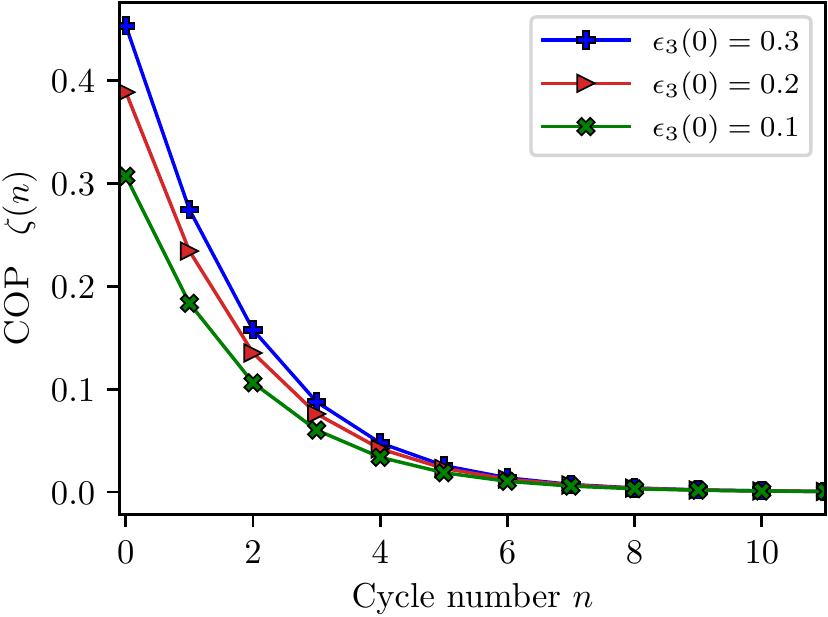}
    \includegraphics{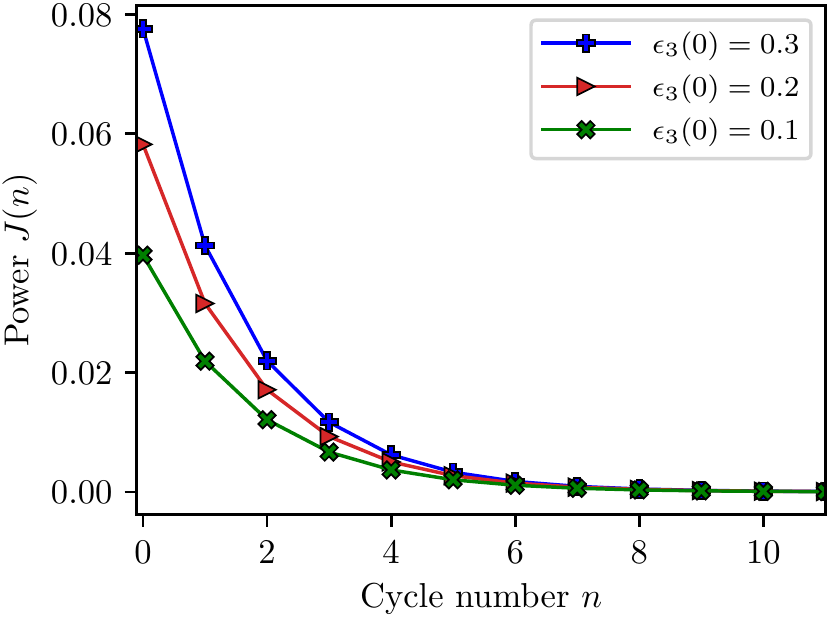}
    \caption{Cooling power \(J(n)\), Eq.~\eqref{15}, and coefficient of performance (COP) \(\zeta(n)\), Eq.~\eqref{17},  as a function of the number \(n\) of cycles, for fixed values of the damping coefficient \(\gamma=0.1\) and of the mixing angle \(\theta=\pi/3\), for different initial polarizations of the reset qubit 3.
    Parameters are \(\epsilon_1(0) = 0\), \(\epsilon_2(0) = 0.3 \).
    }
    \label{fig:diffpol2}
\end{figure}

\section{Theoretical evaluation of the thermodynamic quantities}

We now turn to the evaluation of the thermodynamic quantities of the heat bath algorithmic cooling refrigerator and, in particular, of their fundamental upper bounds in the reversible limit (\(\gamma=0, \theta= \pi/2\)), for arbitrary initial polarizations.
The heat per cycle follows directly from Eq.~\eqref{9} from the definition given in the main text and reads
\begin{equation}
\label{12}
    Q(n) = \big[
        2 ( (1 + \epsilon{2}(0) \epsilon{3}(0)) \epsilon{1}(0) - \epsilon{2}(0) - \epsilon{3}(0) ) \sin^2\theta + \gamma (1 + \epsilon{1}(0)) F(\theta)
    \big]
       \frac{ e^{-nG(\theta,\gamma)} }{4}.
\end{equation}
The corresponding cooling power is  found to be
\begin{equation}
\label{15}
\begin{split}
    J(n) ={}& { Q(n + 1) - Q(n) }
    ={} \left[\frac{(1 - \gamma) F(\theta)}{4} - 1 \right] {Q(n)} \\
     \xrightarrow[\theta = \theta_n]{\gamma \to 0} {}& J_\text{max}(n)=
    [ 1+\epsilon{2}(0) \epsilon{3}(0) ][ \epsilon{2}(0) + \epsilon{3}(0) - \epsilon{1}(0) ( 1 + \epsilon{2}(0) \epsilon{3}(0) ) ] \frac{ e^{- nG(\theta_n, 0) } }{ 2 - 2\epsilon{2}(0) \epsilon{3}(0) } \\
    &+ \gamma \big[2 \epsilon{2}(0)+2 \epsilon{3}(0)+\epsilon{2}(0)^{2} \epsilon{3}(0)^{2}
    - 1 \\
    &\quad- \big( 1+\epsilon{2}(0) \epsilon{3}(0) \big)\big( \epsilon{1}(0) [ n - 3 + (1+n) \epsilon{2}(0) \epsilon{3}(0) ] - n [ \epsilon{2}(0)+\epsilon{3}(0) ] \big)
    \big]
    \frac{ e^{- nG(\theta_n, 0) } }{ 2 - 2\epsilon{2}(0) \epsilon{3}(0) } .
\end{split}
\end{equation}
The cooling power \(J(n)\)  is proportional to the heat \(Q(n)\) since the finite difference of an exponential is again an exponential. The fundamental upper bound \(J_\text{max}(n)\), Eq.~\eqref{15}, generalizes Eq.~(8) of the main text to arbitrary initial polarizations \(\epsilon_i(0)\) of the three qubits. {
   Contrary to  polarization, maximum power is not achieved for  \(\theta \to \pi/2\). This happens because the steady state value \(J(\infty)\) is reached faster  in this limit due to large values of the power in the first two rounds.  This then leads to a suppressed heat removal from the target qubit and a reduced power output. To maximize the refrigeration power \(J(n)\), a solution is to suppress the exponential decay in \(n\) by decreasing the angle \(\theta\) (and therefore the decay coefficient \(G(\theta, \gamma)\)) with the number of strokes. The optimal value of \(\theta\) is
    \begin{equation}
        \theta_n =
        \begin{cases}
            \pi/2 ,\quad& n = 0 \\
            \begin{cases}
                \pi/2 ,\quad& (\epsilon{1}(0), \epsilon{2}(0), \epsilon{3}(0)) \in \mathbb{J} \\
                \theta_\text{opt} ,\quad& \text{otherwise}
            \end{cases} ,\quad& n = 1 \\
            \theta_\text{opt} ,\quad& n \geq 2 ,
        \end{cases}
    \end{equation}
    where the angle \(\theta_\text{opt}\) reads
    \begin{equation}
        \theta_\text{opt} = \frac{1}{2} \arccos\left( \frac{2 \epsilon{2}(0)\epsilon{3}(0) + n \epsilon{2}(0)\epsilon{3}(0) + n - 6}{(2 + n) (1 + \epsilon{2}(0)\epsilon{3}(0))} \right) .
    \end{equation}
   The condition \(\mathbb{J}\) is given by
    \begin{equation}
        \mathbb{J} = \left\{
            (\epsilon{1}(0), \epsilon{2}(0), \epsilon{3}(0))  \in [0, 1]^3
            \ \Big|\
            0 \leq \epsilon{1}(0) < \frac{1}{\sqrt{3}} \cap
            \left(
                0 \leq \epsilon{2}(0) < \frac{1}{3 \epsilon{1}(0)} \cap
                0 \leq \epsilon{3}(0) < \frac{1}{3 \epsilon{2}(0)}
            \right)
        \right\} ,
    \end{equation}
    where we always assumed that \( (\epsilon{2}(0), \epsilon{3}(0)) \geq \epsilon{1}(0) \geq 0 \).
}

On the other hand, using Eq.~\eqref{eq:bath-qubit} and the definition of the work done on the qubit system, we obtain
\begin{multline}
\label{16}
    W(n) = 4 \sin^2\theta \frac{ (1 + \epsilon{2}(0)) (1 + \epsilon{3}(0)) }{(\gamma - 1) F(\theta) + 4}
    + \left(
        1
        + 4 \gamma \sin^2\theta \frac{
            (1 + \epsilon{2}(0) \epsilon{3}(0)) (\gamma - 1)
        }{(\gamma - 1) F(\theta) + 4}
    \right) \\
    \times
    \big(
        2 [ (1 + \epsilon{2}(0) \epsilon{3}(0)) \epsilon{1}(0) - \epsilon{2}(0) - \epsilon{3}(0) ] \sin^2\theta + \gamma (1 + \epsilon{1}(0)) F(\theta)
    \big) \frac{e^{-nG(\theta,\gamma)}}{4} .
\end{multline}
The coefficient of performance eventually follows as
\begin{equation}
\label{17}
\begin{split}
    \zeta(n) ={}& \frac{
       - [ (\gamma - 1) F(\theta) + 4 ] R(\theta, \gamma) e^{-n G(\theta,\gamma)}
    }{
        \big[
            (\gamma - 1) F(\theta) + 4
            - 4 (1 + \epsilon{2}(0) \epsilon{3}(0)) (1 - \gamma) \sin^2(\theta)
        \big] R(\theta, \gamma) e^{-nG }
        + 16 \sin^2(\theta) (1 + \epsilon{2}(0)) (1 + \epsilon{3}(0))
    } \\
    {}&\xrightarrow[\theta = \pi/2]{\gamma \to 0}
   \zeta_\text{max}(n)= 1
    + \frac{ 4 \gamma }{ 1+\epsilon{2}(0) \epsilon{3}(0) }
    \left(
        1 + \frac{
                \left(1+\epsilon{2}(0)\right)\left(1+\epsilon{3}(0)\right)
            }{
                \epsilon{1}(0)-\epsilon{2}(0)-\epsilon{3}(0)+\epsilon{1}(0) \epsilon{2}(0) \epsilon{3}(0)
            } e^{nG(\pi/2, 0) }
    \right)
\end{split}
\end{equation}
where \( R(\theta, \gamma) = 2 \sin^2(\theta) [ (1 + \epsilon{2}\epsilon{3}) \epsilon{1}(0) - \epsilon{2} - \epsilon{3} ] + \gamma (1 + \epsilon{1}(0)) F(\theta) \). The fundamental upper bound \(\zeta_\text{max}(n)\), Eq.~\eqref{17}, generalizes Eq.~(8) of the main text to arbitrary initial polarizations \(\epsilon_i(0)\) of the three qubits.

Heat \(Q(n)\), Eq.~\eqref{12}, and work \(W(n)\), Eq.~\eqref{16} are shown in Fig.~\ref{fig:qw}  as a function of the number \(n\) of cycles, for various values of decay rate \(\gamma\) and of the mixing angle \(\theta\). The influence of  unequal initial polarizations of the reset qubits is illustrated in Figs.~\ref{fig:diffpol} and \ref{fig:diffpol2} for fixed values of the damping rate \(\gamma=0.1\) and of the mixing angle \(\theta=\pi/3\). Work \(W(n)\) and polarization \(\epsilon_1(n)\) of the target qubit, Eq.~\eqref{eq:target-qubit}, are increased when \(\epsilon_3(0) >\epsilon_2(0)\), whereas heat \(Q(n)\) is decreased. At the same time, cooling power \(J(n)\), Eq.~\eqref{15}, and COP \(\zeta(n)\), Eq.~\eqref{17}, are also both increased when \(\epsilon_3(0) >\epsilon_2(0)\).

{\section{Generality of the properties of the imperfect compression gate}}

{The imperfect compression map \(\mathcal{C}\) parametrized by the mixing angle \(\theta\) preserves the target steady state of perfect compression \(\mathcal{C}_{\theta = \pi/2}\) and,  at the same time, slows down the convergence  to the steady state for vanishing dissipation \(\gamma = 0\). We show in this Section that these features generically hold for a family of imperfect compression maps given by a convex combination of two unitaries, namely the ideal compression   and the identity. The structure of these generalized nonideal compression maps will provide additional physical insight into their remarkable properties.}

{Let us consider the trace preserving completely positive map that chooses between the application of the perfect compression gate and the identity (i.e.~do nothing) with a probability distribution \((p_1=\sin^2\theta, p_2=\cos^2\theta)\):
\begin{equation}
\label{21}
    \mathcal{C}_\text{gen}[\bullet] =
    \sin^2\theta \,U \bullet U^\dag + (\bullet) \cos^2\theta,
\end{equation}
where  \(U= \exp(-\mathrm{i} \pi  V/2)\) with \(V=\ketbra{100}{011} + \ketbra{011}{100}\) is the unitary describing ideal compression swap. Such a convex combination of unitary operations (sometimes called random external fields map \cite{ben06}) is unital by construction and thus leaves the maximally mixed state invariant \cite{ben06}. The operator-sum representation of the map \(\mathcal{C}_\text{gen}\) on states diagonal in the energy eigenbasis (and only on those) is identical to that of the compression map \(\mathcal{C}\). The random map \(\mathcal{C}_\text{gen}\) may thus be regarded as a generalization of the compression map \(\mathcal{C}\).}

\begin{figure}[t]
    \centering
    \includegraphics{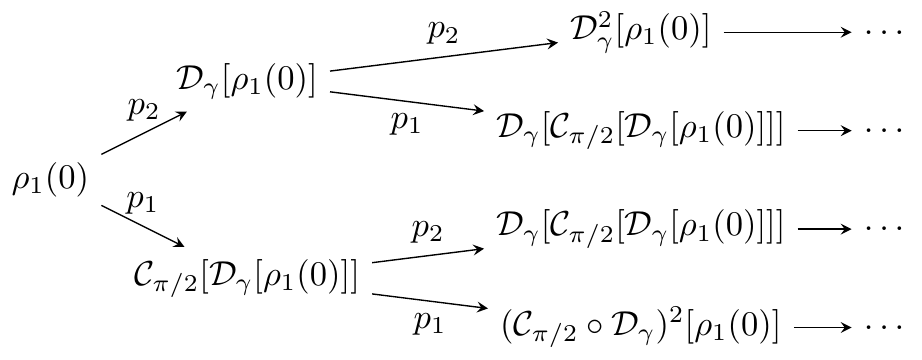}
    \caption{{Graph representing the causal tree of possible evolution paths of fine-grained realizations of the generalized nonideal compression map \eqref{21} given as a convex combination of the ideal compression (with probability \(p_1 = \sin^2\theta\)) and the identity map (with probability \(p_2=\cos^2\theta\)). Nonzero dissipation \(\gamma \neq 0\) leads to different state states of the target qubit
     that depend on the angle \(\theta\). The symbol \(\circ\) indicates composition of maps.}}
    \label{fig:graph}
\end{figure}

{Let us next show that the map \eqref{21} preserve the state state  of the target quit.  In analogy to Eq.~\eqref{eq:full-state}, we introduce the concatenated three-qubit map \(\mathcal{E}_{\theta}=\mathcal{R}[ \mathcal{C}_\text{gen}[ I [\bullet] ] ]\) that combines compression and refresh maps (in the absence of dissipation). We denote by \(\rho_\text{ss}=\mathcal{E}^\infty_{\pi/2}(\rho_0)\) the steady state of  \(\mathcal{E}_{\pi/2}\). We then have
\begin{equation}
\begin{split}
    \lim_{n \to \infty} \mathcal{E}_{\pi/2}^n(\rho_1(0))
    = \lim_{n \to \infty} \mathcal{E}_{\pi/2}(\rho_1(n-1)) ={}& \lim_{n \to \infty}
    \tr_\text{reset}\{
        U (\rho_1(n-1) \otimes \rho_\text{reset}) U^\dag
    \} \\
    ={}& \tr_\text{reset}\{
        U (\rho_\text{ss} \otimes \rho_\text{reset}) U^\dag
    \} \\
    \equiv{}& \rho_\text{ss} .
\end{split}
\end{equation}
Thus, in this limit, the concatenation of dilation, unitary, and trace, acts as an identity operation on \(\rho_\text{ss}\). With this property, we can prove that \(\mathcal{E}_\theta\) has a steady state
which in fact does not depend on \(\theta\):
\begin{equation}
\begin{split}
    \lim_{n \to \infty} \mathcal{E}_\theta(\rho_1(n-1)) ={}& \lim_{n \to \infty}
    \sin^2\theta \tr_\text{reset}\{
        U (\rho_1(n-1) \otimes \rho_\text{reset}) U^\dag
    \}
    + \lim_{n \to \infty} \cos^2(\theta) \rho_1(n-1) \\
    ={}& \sin^2\theta \tr_\text{reset}\{
        U (\rho_\text{ss} \otimes \rho_\text{reset}) U^\dag
    \}
    + \cos^2(\theta) \rho_\text{ss} \\
    ={}& \rho_\text{ss} . \label{eq:ss}
\end{split}
\end{equation}
In summary, at each step the map combines two states sharing the same asymptotic value. As a result, the steady state  \(\rho_\text{ss}\) is preserved.}

{Physically, the angle \(\theta\) interpolates between a map which implements the one-shot ideal compression at every single step (\(\theta = \pi/2\)), and an identity map that  not does nothing (\(\theta = 0\)). In this one-shot regime, the features of the nonideal compression may be intuitively understood:  at every cycle the compression brings the state closer to its stationary value, but in some cycles nothing happens. As a consequence, the steady state is unchanged and the convergence time  increases. This property holds approximately when \(\gamma\) is non-zero, but very small. We also emphasize that this behavior  does not depend on the number of reset qubits.}

{We further note that in  combination with dissipation, the second line in Eq.~\eqref{eq:ss} is no longer valid. The dissipation map \(\mathcal{D}\) not only modifies the overall quantum operation, it also introduces an asymmetry between each realization of the imperfect compression. Consider the graph in Fig.~\ref{fig:graph}: In contrast to the undamped case, each branch in this tree, representing the possible fine-grained paths the system can take, leads to its own steady state. The average will constitute of a typical evolution in this branch, and will thus depend on \(\theta\) to the extent that this typical path depends on the weight of the probability distribution \(p\). For \({\gamma = 0}\), the majority of branches will asymptotically consist of compressions and only a single branch, the uppermost one, consists of only identity operations.}

{\section{Experimental details}}
{This section provides additional details about the experimental implementation of the three-qubit   heat bath algorithmic cooling refrigerator using a system of a NV center in diamond \cite{{zai21}}.}

\begin{figure}[t]
    \centering
    \includegraphics[width=0.54\textwidth]{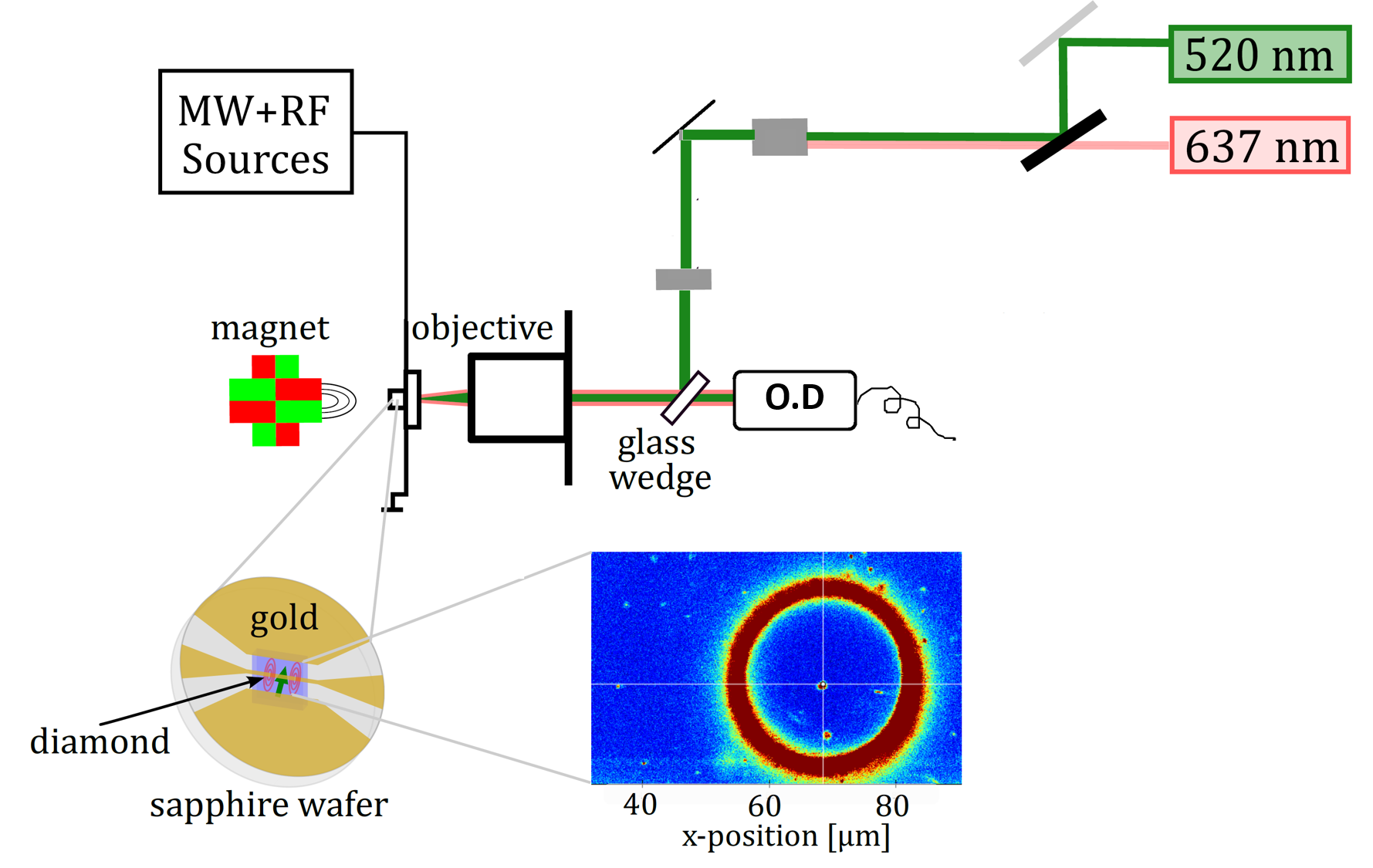}
        \caption{{Sketch of the experimental setup. The setup consists of a homebuilt confocal microscope, a permanent magnet and microwave (MW) and radio frequency
(RF) sources. The 520nm laser is operated at a power close to NV center saturation (0.1mW to 0.5mW before the objective). An additional 637nm laser is used for electron spin repolarization (charge state control) and thus has a power of less than 10 \(\mu\)W. O.D is the standard optical detection setup where the fluorescence is filtered by a 650-nm long-pass filter and a 50-\(\mu\)m pinhole, and then detected by a single-photon-counting avalanche photodiode. At the bottom are shown the sample, substrate and the confocal image displaying the location of the NV center.}}
    \label{fig:polgate}
\end{figure}

{
The experimental setup consists of a homebuilt confocal
microscope, a permanent magnet for the creation of the external magnetic field
and equipment for electron and nuclear spin manipulation as shown in Fig.~\ref{fig:polgate}. The setup operates at ambient conditions, i.e. room temperature and atmospheric pressure and is used exclusively to work with single NV centers. The diamond sample is embedded into a sapphire waver of 2 mm thickness and a  diameter of 50 mm. The sapphire waver is mounted on a 3-axis piezoelectric scanner with a travel range of 100 \(\mu\)m x 100 \(\mu\)m x 25 \(\mu\)m and subnanometer resolution.}

{\subsection{Reset polarization}}
{The SWAP gate used for the reset steps is adapted for the efficient generation of a variable degree of nuclear spin polarization. As compared to the implementation of the traditional SWAP gate using three CNOT gates, here only two CNOT gates are enough.  The final electron spin state after application of the SWAP gate is indeed irrelevant, as it only acts as source of polarization and can be easily repolarized with a green or red laser pulse into \(\ket{m_s= 0}\). Therefore, the third controlled rotation is not required and the SWAP gate simplifies to two controlled spin rotations. Furthermore, to achieve variable polarization transfer to the nuclear spins, the second electron controlled nuclear rotation does not necessarily need to cover the full angle \(\theta=\pi\) but can be replaced by a rotation of variable angle, \(R_{y,\theta}\) as shown in Fig.~\ref{fig:polgate}.}

{For the choice of the magnetic field (\(540\) mT) used in the experiment, direct optical nuclear spin polarization due to GSLAC and ESLAC is not possible as it requires much lower fields (\(\sim 50-100\) mT). The choice for such large fields is to achieve high fidelity single shot readout of the nuclear spins, by improving the nuclear spin life-time that scales quadratically with the field strength \cite{SSR}. The \(^{14}\)N nuclear spin life time reaches close to a ms at such field strengths.}

\begin{figure}[t]
    \centering
    \includegraphics[width=0.59\textwidth]{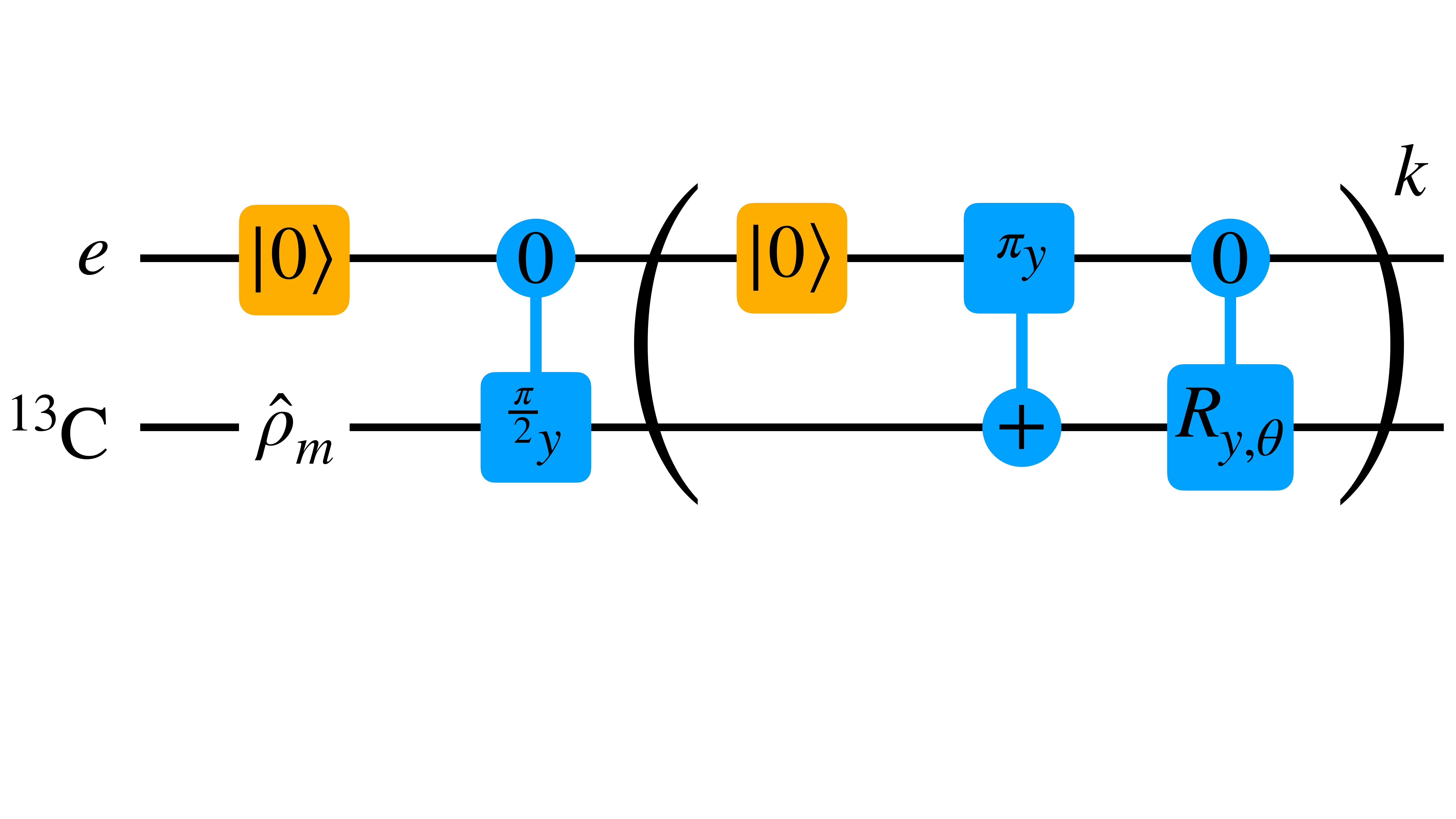}
    \vspace{-15mm}
    \caption{{Pulse sequence for variable degree polarization transfer from electron spin to the two nuclear spins used in the experiment. The
electron initially is in state \(\ket{m_s = 0}\), while the target 13C nuclear spin is in a fully mixed state. To remove any remaining polarization, before the polarization step, a \(50 \mu \)s long \(\frac{\pi}{2}\)-pulse is performed on the nuclear spin. The actual polarization transfer part of the sequence consists of a \(80 \mu\)s red laser pulse for electron reset, a nuclear spin
controlled electron \(\pi\)-pulse (\(6 \mu \text{s}/20 \mu \)s for \({}^{13}C_1/{}^{13}C_2\)) and an electron spin controlled nuclear spin rotation of variable duration (\(0 \mu\)s to \(100 \mu\)s). To increase the nuclear
spin polarization, the polarization transfer part can be repeated
\(k\)-times. Finally, the spin state is read out with single-shot readout (SSR). The experiment was performed for angles \(\theta\) between \(0\) and \(2\pi\).}}
    \label{fig:circuit}
\end{figure}

{\subsection{Gate implementation}}

{The total gate duration of cooling operation \(U\) is \(\sim 284 \mu\)s. An optimal pulse-duration for the nuclear spin gates  was chosen to be around \(50\mu\)s to omit heating of the sample due to the
large RF power and to omit crosstalk to other nuclear spin transitions. The electron spin controlled nuclear spin phase gates do not change the state of the electron spin, thereby avoiding any decoupling errors during the gate operation. Furthermore, the electron spin state remaining in state \(\ket{m_s= 0}\) during the long nuclear spin operations will preserve its coherence over the electron spin relaxation timescales of \(T_{1e}\sim 5.7\) ms. The electron spin \(2\pi\)-pulses take at total duration of \(84 \mu\)s \cite{zai21}.  While the coherences decay on a  timescale of \(T^\text{Hahn}_{2,e} \sim395 \mu\)s. The electron spin gates were optimized with help of the optimal control platform DYNAMO \cite{dynamo} to realize fast and robust Hahn  gates despite electron decoherence on timescales of \(T^\text{Hahn}_{2,e}\) and a dense electron spin spectrum \cite{QEC}.}\\

\section{Experimental determination of the thermodynamic performance}
We evaluate the heat from the set of  qubit polarizations \(\epsilon{1}(n)\) measured using single-shot readout  \cite{zai21}. Using the definition of the heat \(Q(n)\) given in the main text, we concretely have

\begin{equation}
    Q_\text{exp}(n) = - [ \epsilon{1}(n+1) - \epsilon{1}(n) ] .
\end{equation}
The cooling power \(J_\text{exp}(n)\) follows from the finite difference  \(Q_\text{exp}(n+1) - Q_\text{exp}(n)\).

Since the states \(\tilde \rho_i(n)\) of the reset qubits after the compression stage are entirely determined by the target qubit polarization \(\epsilon{1}(n-1)\) via Eq.~\eqref{eq:bath-qubit}, the work \(W(n)\) may be directly evaluated  from the target polarization data (without having to measure the polarizations of the target qubits) and a non-cumulative version of Eq.~\eqref{16}, where the dependence on \(n\) is implicit in using \(\epsilon{1}(n)\) instead of \(\epsilon{1}(0)\). That is,
\begin{equation}
    W_\text{exp}(n) = \sin^2(\theta) \big[ \gamma (\epsilon{2}(0) \epsilon{3}(0) + 1) + (\gamma - 1) (\epsilon{2}(0) \epsilon{3}(0) + 1) \epsilon{1}(n) + \epsilon{2}(0) + \epsilon{3}(0) \big]
    + \epsilon{1}(n) - \epsilon{1}(n+1) .
\end{equation}
We use the experimentally obtained values of \(\epsilon{1}(n)\) to determine the work using the above relation.

\begin{figure}[t]
    \centering
    \includegraphics{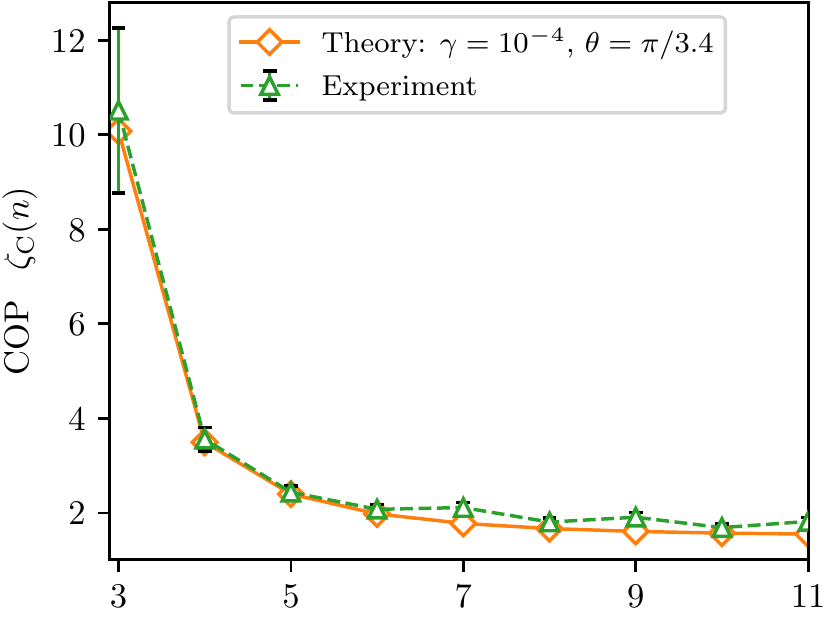}
    \caption{
        Carnot coefficient of performance (COP) \(\zeta_\text{C}(n)\) as a function of the number \(n\) of cycles, for experimentally observed target qubit polarizations, and reset qubit polarization \(\epsilon{2}(0) {\sim} 0.58\).
    }
    \label{fig:expcarnot}
\end{figure}

Using the single-shot readout errors of the three qubits polarizations,
of \({\sim}3\%\) for \(\epsilon{1}(n)\) and \(\epsilon{2}(0)\), and \({\sim}1\%\) for \(\epsilon{3}(0)\), we estimate the relative error of \(W_\text{exp}(n)\) as the standard deviation calculated through conventional linear propagation methods. Likewise, we estimate the relative error of \(Q_\text{exp}(n)\) to be bounded by \({\sim}3\%\) \cite{zai21,qft}. The errors bars for   \(\zeta_\text{exp}(n)\) and \(J_\text{exp}(n)\) follow from the definition of these quantities.

Figure~\ref{fig:expcarnot} additionally shows the   experimental Carnot COP \({\zeta_\text{C}}_\text{exp}(n)\) and the corresponding theoretical expectation \({\zeta_\text{C}}(n)\) for \(\gamma = 10^{-4}\) and \(\theta = \pi/3.4\). The latter are determined uniquely through the temperatures associated to the two polarizations \(\epsilon{1}(n)\) and \(\epsilon{2}(0)\).


\begin{thebibliography}{24}
\bibitem{cal85} H. B. Callen, \textit{Thermodynamics and an Introduction to Thermostatistics}, (Wiley, New York, 1985).
\bibitem{met99} H. J. Metcalf and P. van der Straten, \textit{Laser Cooling and Trapping}, (Springer, berlin, 1999).
\bibitem{let09} V. S. Letokhov, \textit{Laser Control of Atoms and Molecules}, (Oxford University Press, Oxford, 2007).
\bibitem{mac92} P. V. E. McClintock, D. J. Meredit and J. K. Wigmore, \textit{Low-Temperature Physics}, (Springer, Berlin, 1992).
\bibitem{ens05} C. Ens and S. Hunklinger, \textit{Low-Temperature Physics}, (Springer, Berlin, 2005).
\bibitem{nie00} M. A. Nielsen and  I. L. Chuang, \textit{Quantum Computation and Quantum Information}, (Cambridge
University Press, Cambridge, 2000).
\bibitem{des09} E. Desurvire, \textit{Classical and Quantum Information Theory}, (Cambridge University Press, Cambridge, 2009).

\bibitem{sch99} L. J. Schulman and U. V.  Vazirani, Molecular scale heat engines and scalable quantum computation, Proc. 31st ACM Symp. on Theory of Computing, 322 (ACM Press, 1999).
\bibitem{boy02} P. O. Boykin, T.  Mor, V.  Roychowdhury, F.  Vatan, and R. Vrijen, Algorithmic cooling and scalable NMR quantum computers. Proc. Natl Acad. Sci. USA \textbf{99}, 3388 (2002).


\bibitem{par16} D. K. Park, N. A. Rodr\'iguez-Briones, G. Feng, R. R. Darabad, J. Baugh, and R. Laflamme, Heat Bath Algorithmic Cooling with Spins: Review and Prospects, Electron spin resonance (ESR) based quantum computing. Biological Magnetic Resonance. \textbf{31}, 227 (2016).


\bibitem{fer04} J. M. Fernandez, S.  Lloyd, T.  Mor, and V. Roychowdhury,  Algorithmic cooling of spins: a practicable method for increasing polarization, Int. J. Quantum Inf. \textbf{2}, 461 (2004).
\bibitem{sch05} L. J. Schulman, T. Mor, and Y. Weinstein, Physical limits of heat-bath algorithmic cooling, Phys. Rev. Lett. \textbf{94}, 120501 (2005).
\bibitem{sch07} L. J. Schulman, T. Mor, and Y. Weinstein, Physical limits of heat-bath algorithmic cooling, SIAM J. Comput. \textbf{36}, 1729 (2007).
\bibitem{rem07} F. Rempp, M. Michel, and G. Mahler, Phys. Rev. A \textbf{76}, 032325 (2007).
\bibitem{kay07} P. Kaye, Cooling algorithms based on the 3-bit majority, Quantum Inf. Process. \textbf{6}, 295 (2007).
\bibitem{bra14} G. Brassard, Y. Elias, T. Mor, Y. Weinstein, Prospects and Limitations of Algorithmic Cooling,  Eur. Phys. J. Plus \textbf{129}, 258 (2014).

\bibitem{rai15} S. Raeisi and M.  Mosca, Asymptotic bound for heat-bath algorithmic cooling, Phys. Rev. Lett. \textbf{114}, 100404 (2015).
\bibitem{rod16} N. A. Rodr\'iguez-Briones and R. Laflamme, Achievable polarization for heat-bath algorithmic cooling, Phys.
Rev. Lett. \textbf{116}, 170501 (2016).
\bibitem{rai19} S. Raeisi, M. Kieferov,  and M. Mosca,  Novel technique for robust optimal algorithmic cooling, Phys. Rev. Lett. \textbf{122}, 220501 (2019).
\bibitem{rai21} S. Raeisi, No-go theorem behind the limit of the heat-bath algorithmic cooling, Phys. Rev. A \textbf{103}, 062424 (2021).

\bibitem{bau05} J. Baugh, O. Moussa, C. A. Ryan, A.  Nayak, and R.  Laflamme, Experimental implementation of heat-bath algorithmic cooling using solid-state nuclear magnetic resonance, Nature \textbf{438}, 470 (2005).
\bibitem{rya08} C. A. Ryan, O. Moussa, J. Baugh, and R. Laflamme, Spin based heat engine: Demonstration of multiple rounds of algorithmic cooling, Phys. Rev. Lett. \textbf{100}, 140501 (2008).
\bibitem{par15} D. K. Park, G. Feng, R. Rahimi, S. Labruyere, T. Shibata,
S. Nakazawa, K. Sato, T. Takui, R. Laflamme, and J. Baugh, Hyperfine spin qubits in irradiated malonic acid: heat-bath algorithmic cooling,
Quantum Inf. Process. \textbf{14}, 2435 (2015).
\bibitem{ata16} Y. Atia, Y. Elias, T. Mor, and Y. Weinstein, Algorithmic cooling in liquid-state nuclear magnetic resonance, Phys. Rev. A \textbf{93}, 012325 (2016).
\bibitem{zai21} S. Zaiser, C. T. Cheung, S. Yang, D. B. R. Dasari,  S. Raeisi and J. Wrachtrup, Cyclic cooling of quantum systems at the saturation limit, npj Quant. Info. \textbf{7}, 92 (2021).

\bibitem{kos14} R. Kosloff and  A. Levy, Quantum heat engines and refrigerators: Continuous devices, Annu. Rev. Phys. Chem. \textbf{65}, 365 (2014).
\bibitem{rez09} Y. Rezek, P. Salamon, K. H. Hoffmann and R. Kosloff, The quantum refrigerator: The quest for absolute zero, EPL \textbf{85}, 30008 (2009).
\bibitem{all10} A. E. Allahverdyan, K. Hovhannisyan and G. Mahler, Phys. Rev. E \textbf{81}, 051129 (2010).
\bibitem{aba16} O. Abah and E. Lutz, Optimal performance of a quantum Otto refrigerator, EPL \textbf{113}, 60002 (2016).

\bibitem{com1} An information-theoretic analysis of the performance of heat-bath algorithmic cooling, viewed from the perspective of feedback cooling, has been presented in Ref.~\cite{liu16}.
\bibitem{liu16} P. Liuzzo-Scorpo, L. A. Correa, R. Schmidt, and G. Adesso, Thermodynamics of quantum feedback cooling, Entropy \textbf{18}, 48 (2016).

\bibitem{ond81} M. J. Ondrechen, B. Andresen, M. Mozurkewich, and R. S. Berry, Maximum work from a finite reservoir by sequential Carnot cycles, Am. J.  Phys. \textbf{49}, 681 (1981).
\bibitem{wan16} Y. Wang, Optimizing work output for finite-sized heat reservoirs: Beyond linear response, Phys. Rev. E \textbf{93}, 012120 (2016).
\bibitem{taj17} H. Tajima and M. Hayashi, Finite-size effect on optimal efficiency of heat engines, Phys. Rev. E \textbf{96}, 012128 (2017).
\bibitem{poz18} A. Pozas-Kerstjens, E. G. Brown and K. V. Hovhannisyan, A quantum Otto engine with finite heat baths: energy, correlations, and degradation, New J. Phys. \textbf{20}, 043034 (2018).
\bibitem{moh19} M. H. Mohammady and A. Romito,
 Efficiency of a cyclic quantum heat engine with finite-size baths,
Phys. Rev. E \textbf{100}, 012122 (2019).
\bibitem{ma20} Y. H. Ma, Effect of Finite-Size Heat Source's Heat Capacity on the Efficiency of Heat Engine, Entropy \textbf{22}, 1002 (2020).
\bibitem{gya20} J. A. Gyamfi, Fundamentals of quantum mechanics in Liouville space, Eur. J. Phys. \textbf{41}, 063002 (2020).
\bibitem{sup} See Supplemental Material for further details on the mathematical formalism used and details on the experimental implementation \cite{gil09,hor12,ben06,qft,SSR,dynamo,QEC}.
\bibitem{com} An abstract error analysis of heat-bath algorithmic cooling has been performed in Ref.~\cite{kay07}.
\bibitem{com2} The decreasing values of \(\theta_n\) with the cycle number \(n\) reduce the rate of exponential decay {\(g(\gamma,\theta)\)} of the power \(J(n)\) at each step towards the steady state.
\bibitem{doh13}  M. W. Doherty,  N. B. Manson, P. Delaney, F. Jelezko, J. Wrachtrup, L. C. L. Hollenberg, The nitrogen-vacancy colour centre in diamond. Phys. Rep. \textbf{528}, 1 (2013).
\bibitem{coml} {With the Larmor frequency of the target spin (\(^{14}\)N) being \({\sim}1.66\)MHz and a cycle time including  compression and reset steps of about \({\sim}5\)ms, the heat \(Q(n)\) extracted per cycle is on the order of a few neV, and the power \(J(n)\) is on the order of a few \(\mu\)eV/s or \(10^{-26}\)W.}


\bibitem{gil09} A. Gilchrist,  D. R. Terno, and C. J. Wood, Vectorization of quantum operations and its use, arXiv:0911.2539.
\bibitem{hor12} R. A. Horn and C. R. Johnson, \textit{Matrix analysis, (Cambridge University Press, Cambridge, 2012).}

\bibitem{ben06} I. Bengtsson and K. Zyczkowski, \textit{The Geometry of Quantum States, (Cambridge University Press, Cambridge, 2006).}
\bibitem{qft} V. Vorobyov, S. Zaiser, N. Abt, J. Meinel, D. Dasari, P. Neumann and J. Wrachtrup, Quantum Fourier transform for nanoscale quantum sensing, npj Quant. Info. {\bf 7, 124 (2021).}
\bibitem{SSR} P. Neumann J. Beck, M. Steiner, F. Rempp, H. Fedder, P. R. Hemmer, J. Wrachtrup, and F. Jelezko, Single-shot readout of a single nuclear spin, Science {\bf 329, 542 (2010).}
\bibitem{dynamo} S. Machnes, U. Sander, S. J. Glaser, P. de Fouqui\`eres, A. Gruslys, S. Schirmer, and T. Schulte-Herbr\"uggen, Comparing, optimizing, and benchmarking quantum-control algorithms in a unifying programming framework, Phys. Rev. A {\bf 84, 022305 (2011).}
\bibitem{QEC} G. Waldherr, Y. Wang, S. Zaiser, M. Jamali, T. Schulte-Herbr\"uggen, H. Abe, T. Ohshima, J. Isoya, J. F. Du, P. Neumann and J. Wrachtrup,  Quantum error correction in a solid-state hybrid spin register, Nature {\bf 506, 204 (2014).}

\end{thebibliography}
\end{document}